\documentclass[12pt]{article}
\pdfoutput=1
\usepackage[colorlinks,linkcolor=Blue,citecolor=Blue,bookmarks,bookmarksnumbered]{hyperref}
\usepackage[scaled=0.85]{helvet}
\usepackage{amsmath,amssymb,accents,mathrsfs,XoohmE}
\usepackage{graphicx,color}
\graphicspath{{Figures/}}
\usepackage{booktabs}
\usepackage{multirow}
\usepackage{placeins}
\usepackage{amsmath}
\usepackage{subfigure}
\usepackage{diagbox}

\usepackage{XoohmE}

\def\rI{{\rm I}}
\def\rL{{\rm L}}
\def\rR{{\rm R}}

\definecolor{Green}  {rgb}{0.10,0.70,0.10} 
\definecolor{Orange} {rgb}{1.00,0.50,0.15} 
\definecolor{Red}    {rgb}{0.90,0.00,0.12} 
\definecolor{Purple} {rgb}{0.50,0.25,0.55} 
\definecolor{Turque} {rgb}{0.00,0.65,0.85} 
\definecolor{Blue}   {rgb}{0.00,0.00,1.00} 
\definecolor{Magenta}{rgb}{1.00,0.00,1.00} 
\definecolor{Gold}   {rgb}{1.00,0.75,0.25} 
\definecolor{Seaweed}{rgb}{0.01,0.24,0.09} 
\definecolor{Brown}  {rgb}{0.43,0.26,0.32} 
\definecolor{grey1}  {rgb}{0.20,0.20,0.20} 
\definecolor{grey2}  {rgb}{0.40,0.40,0.40} 
\definecolor{grey3}  {rgb}{0.60,0.60,0.60} 
\definecolor{grey4}  {rgb}{0.80,0.80,0.80} 
\definecolor{grey5}  {rgb}{0.90,0.90,0.90} 
\def\C#1#2{{\ifcase#1\or
             \color{Green}\or \color{Orange}\or \color{Red}\or
              \color{Purple}\or \color{Turque}\or \color{Blue}\or
               \color{Magenta}\or \color{Gold}\or \color{Seaweed}\or
                \color{Brown}\or\color{grey1}\or\color{grey2}\or
                 \color{grey3}\else\color{grey4}\fi#2}}

\definecolor{Slate} {rgb}{0.00,0.45,0.55}

\long\def\CMTR#1{\leavevmode\TC{R}{\sf#1}}

\def\rI{{\rm I}}
\def\rJ{{\rm J}}

\def\rL{{\rm L}}
\def\rR{{\rm R}}

\def\fracm#1#2{\hbox{\large{${\frac{{#1}}{{#2}}}$}}}

\def\vCent#1{\vcenter{\hbox{\hss#1\hss}}}
\def\be{\begin{equation}}
\def\ee{\end{equation}}
\newcommand{\bea}{\begin{eqnarray}}
\newcommand{\eea}{\end{eqnarray}}
\newcommand{\ena}{\end{eqnarray}}

\def\pp{{\mathchoice
          {
              \kern 1pt%
              \raise 1pt
              \vbox{\hrule width5pt height0.4pt depth0pt
                    \kern -2pt
                    \hbox{\kern 2.3pt
                          \vrule width0.4pt height6pt depth0pt
                          }
                    \kern -2pt
                    \hrule width5pt height0.4pt depth0pt}%
                    \kern 1pt
           }
            {
              \kern 1pt%
              \raise 1pt
              \vbox{\hrule width4.3pt height0.4pt depth0pt
                    \kern -1.8pt
                    \hbox{\kern 1.95pt
                          \vrule width0.4pt height5.4pt depth0pt
                          }
                    \kern -1.8pt
                    \hrule width4.3pt height0.4pt depth0pt}%
                    \kern 1pt
            }
            {
              \kern 0.5pt%
              \raise 1pt
              \vbox{\hrule width4.0pt height0.3pt depth0pt
                    \kern -1.9pt  
                    \hbox{\kern 1.85pt
                          \vrule width0.3pt height5.7pt depth0pt
                          }
                    \kern -1.9pt
                    \hrule width4.0pt height0.3pt depth0pt}%
                    \kern 0.5pt
            }
            {
              \kern 0.5pt%
              \raise 1pt
              \vbox{\hrule width3.6pt height0.3pt depth0pt
                    \kern -1.5pt
                    \hbox{\kern 1.65pt
                          \vrule width0.3pt height4.5pt depth0pt
                          }
                    \kern -1.5pt
                    \hrule width3.6pt height0.3pt depth0pt}%
                    \kern 0.5pt
            }
        }}

\def\mm{{\mathchoice
                       {
                             \kern 1pt
               \raise 1pt    \vbox{\hrule width5pt height0.4pt depth0pt
                                  \kern 2pt
                                  \hrule width5pt height0.4pt depth0pt}
                             \kern 1pt}
                       {
                            \kern 1pt
               \raise 1pt \vbox{\hrule width4.3pt height0.4pt depth0pt
                                  \kern 1.8pt
                                  \hrule width4.3pt height0.4pt depth0pt}
                             \kern 1pt}
                       {
                            \kern 0.5pt
               \raise 1pt
                            \vbox{\hrule width4.0pt height0.3pt depth0pt
                                  \kern 1.9pt
                                  \hrule width4.0pt height0.3pt depth0pt}
                            \kern 1pt}
                       {
                           \kern 0.5pt
             \raise 1pt  \vbox{\hrule width3.6pt height0.3pt depth0pt
                                  \kern 1.5pt
                                  \hrule width3.6pt height0.3pt depth0pt}
                           \kern 0.5pt}
                       }}

\def\ad{{\kern0.5pt
                   \alpha \kern-5.05pt \raise5.8pt\hbox{$\textstyle.$}\kern
0.5pt}}

\def\bd{{\kern0.5pt
                   \beta \kern-5.05pt \raise5.8pt\hbox{$\textstyle.$}\kern
0.5pt}}

\def\qd{{\kern0.5pt
                   q \kern-5.05pt \raise5.8pt\hbox{$\textstyle.$}\kern
0.5pt}}
\def\Dot#1{{\kern0.5pt
     {#1} \kern-5.05pt \raise5.8pt\hbox{$\textstyle.$}\kern
0.5pt}}

\catcode`@=11
\def\un#1{\relax\ifmmode\@@underline#1\else
        $\@@underline{\hbox{#1}}$\relax\fi}
\catcode`@=12

\def\a{\alpha}
\def\b{\beta}
\def\c{\chi}
\def\d{\delta}
\def\e{\epsilon}

\def\g{\gamma}

\def\i{\iota}
\def\j{\psi}
\def\k{\kappa}
\def\l{\lambda}
\def\m{\mu}
\def\n{\nu}
\def\o{\omega}

\def\r{\rho}
\def\s{\sigma}
\def\t{\tau}

\def\L{\Lambda}
\def\O{\Omega}
\def\P{\Pi}

\def\S{\Sigma}

\def\ca{{\cal A}}

\def\cf{{\cal F}}

\def\dslash{\not{\hbox{\kern-2pt $\partial$}}}
\def\Dslash{\not{\hbox{\kern-4pt $D$}}}
\def\pslash{\not{\hbox{\kern-2.3pt $p$}}}
 \newtoks\slashfraction
 \slashfraction={.13}
 \def\slash#1{\setbox0\hbox{$ #1 $}
 \setbox0\hbox to \the\slashfraction\wd0{\hss \box0}/\box0 }
 
\def\kcr{{\hbox{\ro \char'170}}}                
\def\ktl{{\hbox{\ro \char'170}}}        
\def\ktr{{\hbox{\ro \char'170}}}        
\def\kbl{{\hbox{\ro \char'170}}}        
\def\kbr{{\hbox{\ro \char'170}}}        

\def\plpl{\raise-2pt\hbox{$\raise3pt\hbox{$_+$}\hskip-6.67pt\raise0.0pt
\hbox{$^+$}\hskip 0.01pt$}}
\def\mimi{\raise-2pt\hbox{$\raise3pt\hbox{$_-$}\hskip-6.67pt\raise0.0pt
\hbox{$^-$}\hskip 0.01pt$}} 

\def\bo{{\raise.15ex\hbox{\large$\Box$}}}               
\def\pa{\partial}                                       
\def\TH{{\raise.2ex\hbox{$\displaystyle \bigodot$}\mskip-4.7mu \llap H \;}}
\def\face{{\raise.2ex\hbox{$\displaystyle \bigodot$}\mskip-2.2mu \llap {$\ddot
        \smile$}}}                                      

\def\dt#1{\on{\hbox{\bf .}}{#1}}                
\def\Dot#1{\dt{#1}}

\def\Tilde#1{\widetilde{#1}}                    
\def\Hat#1{\widehat{#1}}                        
\def\leftrightarrowfill{$\mathsurround=0pt \mathord\leftarrow \mkern-6mu
        \cleaders\hbox{$\mkern-2mu \mathord- \mkern-2mu$}\hfill
        \mkern-6mu \mathord\rightarrow$}
\def\dvec#1{\vbox{\ialign{##\crcr
        \leftrightarrowfill\crcr\noalign{\kern-1pt\nointerlineskip}
        $\hfil\displaystyle{#1}\hfil$\crcr}}}           
\def\dt#1{{\buildrel {\hbox{\LARGE .}} \over {#1}}}     

\def\fracm#1#2{\hbox{\large{${\frac{{#1}}{{#2}}}$}}}
\def\sfrac#1#2{{\vphantom1\smash{\lower.5ex\hbox{\small$#1$}}\over
        \vphantom1\smash{\raise.4ex\hbox{\small$#2$}}}} 
\def\bfrac#1#2{{\vphantom1\smash{\lower.5ex\hbox{$#1$}}\over
        \vphantom1\smash{\raise.3ex\hbox{$#2$}}}}       
\def\afrac#1#2{{\vphantom1\smash{\lower.5ex\hbox{$#1$}}\over#2}}    

\def\pa{\partial}      
\let\bm\relax
\newcommand{\bm}[1]{{\boldsymbol{#1}}}

\def\ad{{\dot{\alpha}}}
\def\bd{{\dot{\beta}}}

 \font\rOpe=cmsy10                        
 \def\ktl{{\hbox{\rOpe\char'170}}}        
 \def\kbl{{\hbox{\rOpe\char'170}}}        
 \def\kcr{{\reflectbox{\rOpe\char'170}}}        
 \def\ktr{{\reflectbox{\rOpe\char'170}}}        
 \def\kbr{{\reflectbox{\rOpe\char'170}}}        
 \def\Border{\vbox{\hsize0pt
        \setlength{\unitlength}{1mm}
        \newcount\xco
        \newcount\yco
        \xco=-21
        \yco=12
        \begin{picture}(0,0)(-7.5,0)
        \put(\xco,\yco){$\ktl$}
        \advance\yco by-1
        {\loop
        \put(\xco,\yco){$\kcr$}
        \advance\yco by-2
        \ifnum\yco \rangle-240
        \repeat
        \put(\xco,\yco){$\kbl$}}
        \xco=170
        \yco=12
        \put(\xco,\yco){$\ktr$}
        \advance\yco by-1
        {\loop
        \put(\xco,\yco){$\kcr$}
        \advance\yco by-2
        \ifnum\yco \rangle-240
        \repeat
        \put(\xco,\yco){$\kbr$}}
        \put(-19.5,13){\scalebox{.6065}{%
         University of Maryland Center for String and Particle  Theory \&\ Physics Department%
        |University of Maryland Center for String and Particle  Theory \&\ Physics Department}}
        \put(-19.5,-241.5){\scalebox{.5835}{%
         ****University of Maryland * Center for String and
         Particle  Theory* Physics Department****University of Maryland *Center
        for String and Particle  Theory* Physics Department}}
        \end{picture}
        \par\vskip-8mm}}
\definecolor{UMred}{rgb}{.9,.05,.2}
\definecolor{HUblue}{rgb}{.0,.3,.7}

\definecolor{Red}    {rgb}{0.90,0.00,0.12} 
\definecolor{Blue}   {rgb}{0.00,0.00,1.00} 
\definecolor{Green}  {rgb}{0.10,0.70,0.10} 
\definecolor{Turque} {rgb}{0.00,0.65,0.85} 
\definecolor{Orange} {rgb}{1.00,0.50,0.15} 
\definecolor{Magenta}{rgb}{1.00,0.00,1.00} 
\definecolor{Gold}   {rgb}{1.00,0.75,0.25} 
\definecolor{Seaweed}{rgb}{0.01,0.24,0.09} 
\definecolor{Purple} {rgb}{0.50,0.25,0.55} 
\definecolor{Brown}  {rgb}{0.43,0.26,0.32} 
\definecolor{grey1}  {rgb}{0.20,0.20,0.20} 
\definecolor{grey2}  {rgb}{0.40,0.40,0.40} 
\definecolor{grey3}  {rgb}{0.60,0.60,0.60} 
\definecolor{grey4}  {rgb}{0.80,0.80,0.80} 
\definecolor{grey5}  {rgb}{0.90,0.90,0.90} 
\def\C#1#2{{\ifcase#1\or
             \color{Red}\or \color{Green}\or \color{Blue}\or\
              \color{Turque}\or \color{Orange}\or \color{Magenta}\or 
               \color{Gold}\or \color{Seaweed}\or \color{Purple}\or
                \color{Brown}\or\color{grey1}\or\color{grey2}\or
                 \color{grey3}\else\color{grey4}\fi#2}}

\definecolor{Slate} {rgb}{0.00,0.45,0.55}

\newdimen\parshift\parshift=\parindent
\catcode`@=11
 \long\def\@footnotetext#1{\insert\footins{\reset@font\footnotesize
           \interlinepenalty\interfootnotelinepenalty\splittopskip%
            \footnotesep\splitmaxdepth\dp\strutbox\floatingpenalty\@MM%
             \hsize\columnwidth\addtolength{\hsize}{-2\parindent}
              \@parboxrestore\protected@edef\@currentlabel%
              {\csname p@footnote\endcsname\@thefnmark}%
                \color@begingroup%
                 \@makefntext{\rule\z@\footnotesep\ignorespaces#1%
                  \@finalstrut\strutbox}%
                \color@endgroup}}
 \long\def\@makefntext#1{\hglue\parshift%
           \vbox{\noindent\baselineskip=11pt plus.5pt minus.5pt\hb@xt@0em{\hss\@makefnmark\kern1pt}#1}}
\catcode`@=12

\newskip\humongous \humongous=0pt plus 1000pt minus 1000pt
\def\caja{\mathsurround=0pt}
\def\eqalign#1{\,\vcenter{\openup2\jot \caja
        \ialign{\strut \hfil$\displaystyle{##}$&$
        \displaystyle{{}##}$\hfil\crcr#1\crcr}}\,}
\newif\ifdtup

\makeatletter
\def\section{\@startsection{section}{1}{\z@}
        {3ex plus-1ex minus-.2ex}{1pt plus1pt}{\large\sf\bfseries\boldmath}}
\def\subsection{\@startsection{subsection}{2}{\z@}
         {1.5ex plus-1ex minus-.2ex}{0.01pt plus1pt}{\sf\slshape}}
\def\subsubsection{\@startsection{subsubsection}{3}{\z@}
          {1.5ex plus-1ex minus-.2ex}{0.01pt plus0.2pt}{\sf\boldmath}}
\def\paragraph{\@startsection{paragraph}{4}{\z@}
           {.75ex \@plus.5ex \@minus.2ex}{-2mm}{\sf\bfseries\boldmath}}
\makeatother

 \allowdisplaybreaks
 \seceq

\usepackage{lipsum}
\usepackage{listings}
\definecolor{MyDarkGreen}{rgb}{0.0,0.4,0.0} 
\def\VV{$ 
\eqalign{
&({\cal P}{}_{[1]}|{\cal P}{}_{[3]})  ({\cal P}{}_{[2]}|{\cal P}{}_{[5]}) \cr 
&({\cal P}{}_{[4]}|{\cal P}{}_{[6]})  
}$}

\def\WW{$
({\cal P}{}_{[2]}|{\cal P}{}_{[3]})$}

\def\XX{$ 
\eqalign{
&({\cal P}{}_{[1]}|{\cal P}{}_{[2]})  ({\cal P}{}_{[1]}|{\cal P}{}_{[6]}) \cr
&({\cal P}{}_{[2]}|{\cal P}{}_{[6]})  ({\cal P}{}_{[3]}|{\cal P}{}_{[4]}) \cr
&({\cal P}{}_{[3]}|{\cal P}{}_{[5]})  ({\cal P}{}_{[4]}|{\cal P}{}_{[5]})
}$}

\def\YY{$\eqalign{
({\cal P}{}_{[1]}|{\cal P}{}_{[5]}) & ({\cal P}{}_{[2]}|{\cal P}{}_{[4]}) \\ 
({\cal P}{}_{[3]}|{\cal P}{}_{[6]})  & {~}
}$}

\def\ZZ{$
({\cal P}{}_{[1]}|{\cal P}{}_{[4]}) ({\cal P}{}_{[5]}|{\cal P}{}_{[6]}) 
$}

\usepackage[enableskew,vcentermath]{youngtab}
\let\TC=\textcolor
\definecolor{Hey}{rgb}{.9,.05,.4}
\definecolor{orange}{rgb}{1,.5,0}
\definecolor{plum}{rgb}{.4,0,.6}
\definecolor{R}{rgb}{1,0,0}
\definecolor{G}{rgb}{0.1,0.7,0}
\definecolor{B}{rgb}{0,0,1}

\long\def\CMTR#1{\leavevmode\TC{R}{\sf#1}}

\begin{document}

\thispagestyle{empty}
\noindent{\small
\hfill{  \\ 
$~~~~~~~~~~~~~~~~~~~~~~~~~~~~~~~~~~~~~~~~~~~~~~~~~~~~~~~~~~~~~~~~~$
$~~~~~~~~~~~~~~~~~~~~~~~~~~~~~~~~~~~~~~~~~~~~~~~~~~~~~~~~~~~~~~~~~$
{}
}}
\vspace*{0mm}
\begin{center}
{\large \bf
The 300 ``Correlators" Suggests 4D, $\cal N $ = 1 SUSY\\[2pt]
Is a Solution to a Set of Sudoku Puzzles \\[2pt]
}   \vskip0.3in
{\large {
$~~~~~~~~~~~~~$
Aleksander J.\ Cianciara\footnote{aleksander${}_-$cianciara@brown.edu}$^{a,b}$,
S.\ James Gates, Jr.\footnote{sylvester${}_-$gates@brown.edu}$^{,a,b}$, 
\newline
$~~~~~~~~~~~~$
Yangrui Hu\footnote{yangrui$_-$hu@brown.edu}${}^{,a,b}$, and
Renée Kirk\footnote{zkirk@terpmail.umd.edu}$^{,c}$ $~~~~~~$
}}
\\*[8mm]
\emph{
\centering
$^{a}$Brown University, Department of Physics,
\\[1pt]
Box 1843, 182 Hope Street, Barus \& Holley 545,
Providence, RI 02912, USA,
\\[12pt]
$^{b}$Brown Center for Theoretical Physics, 
\\[1pt]
340 Brook Street, Barus Hall,
Providence, RI 02912, USA,
\\[4pt] and \\[4pt]
$^{c} $Department of Physics, University of Maryland,
\\[1pt]
College Park, MD 20742-4111, USA
}
 \\*[20mm]
{ ABSTRACT}\\[4mm]
\parbox{142mm}{\parindent=2pc\indent\baselineskip=14pt plus1pt
A conjecture is made that the weight space for 4D, $\cal N$-extended supersymmetrical 
representations is embedded within the permutahedra associated with permutation groups 
${\mathbb{S}}{}_{d}$. Adinkras and Coxeter 
Groups associated with minimal representations of 4D,
$\cal N$ = 1 supersymmetry provide evidence supporting this conjecture. It is shown that the 
appearance of the mathematics of 4D, $\cal N$ = 1 minimal off-shell supersymmetry representations
is equivalent to solving a four color problem on the truncated octahedron.  This observation 
suggest an entirely new way to approach the off-shell SUSY auxiliary field problem based
on IT algorithms probing the properties of ${\mathbb{S}}{}_{d}$.}
 \end{center}
\vfill
\vfill
\clearpage
%

%
\section{Introduction}
\label{sec:NTR0}

A research work \cite{permutadnk} in 2014 revealed a previously unsuspected link between
$\cal N$ = 1 supersymmetrical field theories in four dimensional spacetimes and Coxeter Groups
\cite{Cx[1],Cx[2],Cx[3]}.  Prior to the 2014 work, a program had been created whereby
4D, $\cal N$ = 1 SUSY theories had been reduced to 1D, $\cal N$ = 4 SUSY theories and
there was shown the ubiquitous appearance of certain types of matrices \cite{GRana1,GRana2}
that eventually were identified with an elaboration of adjacency matrices for a class
of graphs given the monikers of ``adinkras'' \cite{Adnk1}. An even later development occurred
with the realization \cite{adnkGEO1,adnkGEO2} that these graphs correspond to a special class of 
Grothendieck's  ``dessins d'enfant" in 
algebraic geometry.

These modified adjacency matrices
were found to occur in two types indicated by the symbols ${\bm {\rL}}{}_{{}_{\rI}}$
and ${\bm {\rR}}{}_{{}_{\rI}}$.  The first type is associated with the transformations
of bosons into fermions under the action of supercharges, while the second type is 
associated with the transformations of fermions into bosons under the action of supercharges.

For 4D, $\cal N$ = 1 off-shell theories, the index 
$\rI$ takes on values 1, $\dots $, 4 and the matrices are real 4p $\times$ 4p matrices 
for some integer $p$.  The minimal non-trivial representations occur for $p$ = 1.
These real matrices satisfy an algebra given by
\begin{align}
\begin{split}
{\bm {\rL}}{}_{{}_{\rI}} \, {\bm {\rR}}{}_{{}_{\rJ}} ~+~ {\bm {\rL}}{}_{{}_{\rJ}} \, {\bm {\rR}}{}_{{}_{\rI}}
& ~=~ 2\,  \d{}_{{}_{\rm {I \, J}}} \, {\bm {\mathbb{I}}}{}_4  ~~~, ~~~
{\bm {\rR}}{}_{{}_{\rI}} \, {\bm {\rL}}{}_{{}_{\rJ}} ~+~ {\bm {\rR}}{}_{{}_{\rJ}} \, {\bm {\rL}}{}_{{}_{\rI}}
~=~ 2\,  \d{}_{{}_{\rm {I \, J}}} \, {\bm {\mathbb{I}}}{}_4 ~~~.
\end{split}
\label{eq:GAlg1}
\end{align}
and which is referred to as ``the Garden Algebra.'' Some years \cite{ENUF} prior to this suggested name, 
it was noted  these conditions are closely related to the definition of a special class of real
Euclidean Clifford algebras.  The real matrices ${\bm {\rL}}{}_{{}_{\rI}}$ and ${\bm {\rR}}{}_{{}_{\rI}}$
can be used to form $8p$ $\times$ $8p$ matrices using the definition
\be{
{\Hat {\bm \g}}{}_\rI ~=~ \left[\begin{array}{cc}
~0 & ~~  {\bm {\rL}}_\rI  \\
{}~&~\\
~ {\bm {\rR}}_\rI & ~~ 0 \\
\end{array}\right]  ~~~,
} \label{CLFF} \ee
which form a Clifford Algebra, with respect to the Euclidean metric defined by the Kronecker delta symbol
$ \delta_{\rm {IJ}}$
\begin{equation}
\left\{{\Hat {\bm \g}}{}_\rI, {\Hat {\bm \g}}{}_\rJ \right\} = 2 \delta_{\rm {IJ}} \, {\bm {\mathbb{I}}}{}_8 
~~~.
\label{CLFF2}
\end{equation}
Here the symbol ${\bm {\mathbb{I}}}{}_8$ denotes the 8 $ \times$ 8 identity matrix to be distinguished from the
case of the 4$ \times$4 identity matrix ${\bm {\mathbb{I}}}{}_4$. 

There exists ten distinct minimal off-shell 4D, $\cal N$ = 1 supermultiplets as indicated below:
\newline
\noindent 
$~~~~~~~~$ (S01.) $ {\rm {Chiral~Supermultiplet:}} ~{(A, \, B, \,  \psi_a , \, F, \, G)} ~~~,$
$~$ \newline \noindent 
$~~~~~~~~$ (S02.) $ {\rm {Hodge-Dual~ \#1~Chiral~Supermultiplet:}} ~{({\widehat A}, \, {\widehat B}, \,  \psi_a , \, {\rm f}_{\mu
 \, \nu \, \rho}, \, {\widehat G})}  ~~~, $
$~$ \newline \noindent 
$~~~~~~~~$ (S03.) ${\rm {Hodge-Dual~ \#2~Chiral~Supermultiplet:}} ~{({\Tilde A}, \, {\Tilde B}, \,  \psi_a , \, {\widehat F}, \, {\rm 
g}_{\m \, \n \, \rho})} ~~~,$
$~$ \newline \noindent 
$~~~~~~~~$ (S04.) $ {\rm {Hodge-Dual~ \#3~Chiral~Supermultiplet:}} ~{(\check{A}, \, \check{B}, \,  \psi_a , \, 
{\check {\rm f}}_{\m \, \n \, \rho}, \, {\check {\rm g}}_{\m \, \n \, \rho})}~~~, $
\newline \noindent 
$~~~~~~~~$ (S05.) $ {\rm {Tensor~Supermultiplet:}} ~{(\varphi, \, B{}_{\mu \, \nu }, \,  \chi_a )}   ~~~,$
$~$ \newline \noindent 
$~~~~~~~~$ (S06.) $ {\rm {Axial-Tensor~Supermultiplet:}} ~{({\widehat {\varphi}}, \, {\widehat B}{}_{\mu \, \nu }, \,  {\widehat {\chi}}_a )} ~~~,$
$~$ \newline \noindent 
$~~~~~~~~$ (S07.) $ {\rm {Vector~Supermultiplet:}}~ (A{}_{\mu} , \, \l_b , \,  {\rm d}) ~~~,$
$~$ \newline \noindent 
$~~~~~~~~$ (S08.) ${\rm {Axial-Vector~Supermultiplet:}}~ (U{}_{\mu} , \, {\widehat \l}_b , \,  {\widehat {\rm d}
)}~~~, $
$~$ \newline \noindent 
$~~~~~~~~$ (S09.) ${\rm {Hodge-Dual~Vector~Supermultiplet:}}~ ({\Tilde A}{}_{\mu} , \, {\Tilde \l}_b , \, {\Tilde {\rm d}}
{}_{\mu \, \nu \, \rho} )~~~,$
$~$ \newline \noindent 
$~~~~~~~~$ (S10.) ${\rm {Hodge-Dual~ Axial-Vector~Supermultiplet:}}~ ({\breve U}{}_{\mu} , \, {\breve \l}
{}_b , \,  {\breve {\rm d}}{}_{\mu \, \nu \, \rho} ) ~~~.$
$~$  \newline $~$
\newline
\noindent
Each of these systems can be reduced to one dimension by simply ignoring the spatial dependence of the
field variables.  If we ignore the supermultiplets that are equivalent to one another via 3-form Hodge duality\footnote{In the works of \cite{HYMN1,HYMN2}
a discussion of the use of eigenvalues determined by ${\bm {\rL}}_\rI$ and ${\bm {\rR}}_\rI$ matrices has been presented to describe the relation among such Hodge 1-form/3-form dual supermultiplets.}or 
parity transformations, this leaves only the supermultiplets described by the fields contained in (S01.), (S05.), and (S07.).  

The work in \cite{permutadnk} showed that each of these remaining three supermultiplets possess distinct 
${\bm {\rL}}{}_{{}_{\rI}}$ and ${\bm {\rR}}{}_{{}_{\rI}}$ matrices that can be factorized.  However
this work, (based on a search algorithm), explicitly constructed many more such sets of matrices.
The twelve matrices obtained from the reduction of the three supermultiplets were shown to be only 
a small number of 384 matrices that satisfy the ``Garden Algebra'' in quartets.  For this larger number, explicit 
expressions of the form
(there is no sum over the $ {\, \widehat {\rI}}$ index implied on the RHS of this equation),
\begin{equation}
 {\bm {\rL}}_{\, \widehat {\rI}} 
 ~=~  {\bm {\cal S}}_{\, \widehat {\rI}}  \, {\bm {\cal P}}_{\, \widehat {\rI}}  ~~~.
\label{eq:aas1}
\end{equation}
were shown for every matrix.  Thus, the matrices found by the search algorithm constitute 96 quartets of matrices 
that satisfy the condition shown in Eq.\ (\ref{CLFF2}).
Here we use the index $ {\, \widehat {\rI}}$ that takes on values 1, 2, $\dots$, 384 in order to denote
each of these matrices.  Moreover, fixing the value of $ {\, \widehat {\rI}}$, the matrices $  {\bm {\cal S}}_{\, \widehat {\rI}} $
were all found to take the form of purely diagonal matrices that each square to the identity matrix.  We should be clear that the entirety of 
these 384 matrices do {\em {not}} form a Clifford algebra.  Instead, judicious choices of quartets of these
do satisfy the conditions shown in Eq.\ (\ref{CLFF2}).

In a very similar manner, for a fixed value of $ {\, \widehat {\rI}}$, all of the matrices $  {\bm {\cal P}}_{\, \widehat {\rI}} $,
take the forms of elements in the permutation group (${\bm {\mathbb{S}}}{}_4$) of order four.  However, in order to satisfy 
the condition in Eq. (\ref{CLFF2}), the entirety of the permutation group was found to be dissected into six distinct 
{\textit{unordered}} subsets.  One of the main purposes of this current work is to revisit this dissection behavior
and discover how this can be uncovered without the use of supersymmetry-based arguments, but only relying on properties of
${\bm {\mathbb{S}}}{}_{4}$.  This approach will lead to a well-known algebraic polytope associated with the permutation groups,
the permutahedron.  The permutahedron is a polytope whose vertices represent the elements of the permutations acting on the first 
$\cal N \,-$ 1  integers. 
For $\cal N$ = 1 there are 24 elements in the set of permutations which means that the factor of the 
${\bm {\cal S}}_{\, \widehat {\rI}} $ matrices that appear in eq.\ (\ref{eq:aas1}) contains sixteen copies of the
permutation elements.

The dissection can be seen from this perspective to arise from the symmetries of the permutahedron and the use of weak Bruhat ordering.

The current work is organized as follows.

Chapter two is dedicated to an explanatory discussion of the two sets of notational conventions we use to describe elements of the permutation group.  The description of how to construct a ``dictionary'' 
(which is actually given in chapter three) is provided for the ``translation'' between the two sets of conventions.

Chapter three contains a review of the previous discussion of how the use of projection techniques from 4D, $\cal N$ = 1
SUSY to 1D, $\cal N$ = 4 SUSY implied a dissection of ${\mathbb S}{}_4$ was required to be consistent with SUSY.  Venn diagrams are used to explain how
``Kevin's Pizza'' (a dissection of ${\mathbb S}{}_4$ in six disjoint subsets) was discovered as a result of projecting 
4D, $\cal N$ = 1 SUSY to 1D, $\cal N$ = 4 on theories that are related by Hodge Duality in four dimensions.

Chapter four provides a concise and simplified discussion of Bruhat weak ordering and how (when applied to ${\mathbb S}{}_4$) it acts to create a ``distance measuring function" of the elements of the permutation group.  This is
contrasted with the Hamming distance that is defined on strings of bits.  It is noted
that the truncated octahedron provides a framework for being overlaid by a network created by
the Bruhat weak ordering to lead
to the permutahedron.

Chapter five introduces the word ``correlators'' to define the minimal
path length between any number of the vertices of the permutahedron.  The focus
of the presentation is only upon two point correlators, though it's easily possible to define $n$-point correlators by analogy.  An investigation of correlators for
permutations from SUSY quartets is performed.  The complete results for the 
two-point correlators associated with these are reported.  This is the first part of studying all three hundred two-point correlators.

Chapter six contains the calculations of two-point correlators among distinct
pairs of the quartets. This completes the calculation of the three hundred two-point correlators. In particular, it is found that
the quartets of permutation generators that are consistent with SUSY are those
with the property that the sums of their eigenvalues exceed non-quartet calculations
when compared to pairs of distinct quartets.

Chapter seven contains sets of calculations similar to those of the preceding two
chapters.  However, the emphasis here is upon the study of such correlators among
the chromotopology of four color adinkras.  The values of twenty thousand seven hundred and thirty-six in terms of eigenvalues and traces are reported.

Chapter eight sets the stage for future investigations.  The permutation generators for {\em {all}} values of $\cal N$ were established in papers
\cite{GRana1,GRana2} written in the 1995-1996 period and it is noted that this implies
the opportunity to study the correlators with $\cal N$ greater than four ( the focus of this work).

A final chapter containing the conclusion, one appendix containing explicit matrix definitions of
${\mathbb S}{}_4$ permutation elements, a second appendix reporting on
the correlators associated with the chromotopologies,
and a bibliography close out the discussion.

\newpage
\section{Notational Conventions}
\label{sec:DeFNs1}

In Appendix \ref{appen:Pmatrices}, an explicit matrix representation of all twenty-four elements of 
${\mathbb{S}}{}_{4}$
is given.  However, in order to more efficiently give the presentation, there are two notational conventions introduced to avoid explicitly writing the matrices.

There exists what may be called the ``parenthetical convention,'' or ``cycle convention.''  In this convention, parenthesis marks $()$ surround a set of numerical entries ``$a_1$, $a_2$, $\dots $", e.\ g.\ $(a_1 a_2 a_3 \, \dots)$ which provides an instruction that $a_1$ $\to$ $a_2$, $a_2$ $\to$ $a_3$, $a_3$ $\to$ $a_4$,
etc. until one gets to the last element
which is instructed 
to ``move'' to the position of the first element.  This describes the cycles that are induced by any permutation element.  The symbol ``$()$" implies no replacements or ``movements'' and hence denotes the identity permutation element.  This notational convention also eskews ``one-cycles'' (i.\ e.\ $(a)$) since these are equivalent to the identity permutation also.
Thus, the number of elements enclosed is either zero, two, three, or four. It should be noted this is a ``unidirectional" convention where the
symbol is to be read from left to right.
Finally, the maximum number of distinct elements that can appear is four.  In
Appendix \ref{appen:Pmatrices} this notation is related to the explicit matrix representation.

There is also what can be called the ``bra-ket convention.''  In this convention,
the ``bra-ket'' symbol  $ \langle \rangle$ is used to always enclose four elements
 $\langle a_1 a_2 a_3 a_4 \rangle$. Perhaps the most efficient manner to present this is via four examples as shown in eq. (\ref{equ:DcT}) 
 
\begin{equation}
    \begin{split}
        \begin{matrix}
            \langle 1234 \rangle\\
            ()
        \end{matrix}~&\Rightarrow~ 
        \begin{matrix}
            \langle \CMTR{1234} \rangle\\
             \langle 1234 \rangle
        \end{matrix}~\Rightarrow~
         1\rightarrow1,\, 2\rightarrow2,\, 3\rightarrow3,\, 4\rightarrow4,\\
         \begin{matrix}
            \langle 1324 \rangle\\
            (23)
        \end{matrix}~&\Rightarrow~ 
        \begin{matrix}
            \langle \CMTR{1234} \rangle\\
             \langle 1324 \rangle
        \end{matrix}~\Rightarrow~
         1\rightarrow1,\, 2\rightarrow3,\, 3\rightarrow2,\, 4\rightarrow4,\\
           \begin{matrix}
            \langle 1342 \rangle\\
            (234)
        \end{matrix}~&\Rightarrow~ 
        \begin{matrix}
            \langle \CMTR{1234} \rangle\\
             \langle 1342 \rangle
        \end{matrix}~\Rightarrow~
         1\rightarrow1,\, 2\rightarrow3,\, 3\rightarrow4,\, 4\rightarrow2,\\
           \begin{matrix}
            \langle 2143 \rangle\\
            (12)(34)
        \end{matrix}~&\Rightarrow~ 
        \begin{matrix}
            \langle \CMTR{1234} \rangle\\
             \langle 2143 \rangle
        \end{matrix}~\Rightarrow~
         1\rightarrow2,\, 2\rightarrow1,\, 3\rightarrow4,\, 4\rightarrow3,\\
    \end{split}
    \label{equ:DcT}
\end{equation}


In the leftmost column the ``bra-ket'' convention is written above its 
equivalent ``parenthetical" equivalent.  To the immediate right of the
``bra-ket" description and in red numbers appears the fiducial bra-ket
description of the identity element.  By comparing the entries in the red 
numbered fiducial bra-ket to the black lettered entries immediately below
it, one can record the actions of the ``bra-keted'' element on the numbers
1, 2, 3, and 4 and easily extract the cycles.  The cycles are then recorded
into the ``parenthetical'' description at the bottom left most entries in
the first column for each example.  Thus, the translations can be read
vertically by looking at the leftmost column for each example.

One other convention that provides a foundation of our exploration is to
use lexicographical ordering of the permutation elements.

For the bra-ket convention, the lexicographical order is simple to ``read
off.'' Given the permutation element $\langle a_1 a_2 a_3 a_4 \rangle$ we simply map 
this to the numerical value of $a_1$, $a_2 a_3 a_4$.  For two or more
permutation elements, one uses the mapping and the usual magnitude
of the numerical equivalents to determine the lexicographical order.
One example of the mapping is provided by $\langle 1 2 3 4 \rangle$ $\to$ 1,234 and
another is provided by $\langle 2 3 1 4  \rangle$ $\to$ 2,314.  So the lexicographical 
ordering of $\langle 2 3 1 4  \rangle$ and $\langle 1 2 3 4  \rangle$ is  $\langle 1 2 3 4  \rangle$ first 
followed by $\langle 2 3 1 4  \rangle$.

If one were to attempt to extend the concept
of lexicographical ordering into the realm of the parenthetical 
basis, an unacceptable degree of arbitrariness enters.   To see
this a simple example suffices.  

Consider the two distinct permutations which in the $()$-basis are expressed as 
$(12)(34)$ and $(1234)$.  The first has two distinct 2-cycles and the second has 
a single 4-cycle. While it is possible to introduce some additional rules over and 
above that which is used in the $\langle \rangle$-basis, this would also introduce arbitrariness.

\newpage
\section{Dissecting A Coxeter Group Into `Kevin's Pizza'}
\label{sec:DeFNs2}

The 384 elements of the Coxeter Group ${\mathbb{BC}}{}_{4}$ are represented in Eq.\ (\ref{eq:aas1}) and
upon taking the absolute values of the matrix representations, one obtains sixteen copies of ${\mathbb{S}}{}_{4}$,
the group of permutations of four objects.

Lexicographically listed in the $\langle \rangle$-basis, the twenty-four permutation elements of ${\mathbb{S}}{}_{4}$ are
given by
\be   \eqalign{
\{ {\mathbb{S}}{}_{4} \} ~&=~ \{ ~ \langle 1234 \rangle, \,   \langle 1243 \rangle, \, \langle 1324 \rangle,  \,     \langle 1342 \rangle, \,   \langle 1423 \rangle, \,  \langle 1432 \rangle, \cr
&{}~{~~~\,~~~}  \langle 2134 \rangle,  \,  \langle 2143 \rangle , \,  \langle 2314 \rangle,  \,   \langle 2341 \rangle,  \,    \langle 2413 \rangle,  \,   \langle 2431 \rangle, 
 \cr
&{}~{~~~\,~~~} \langle 3124 \rangle, \, \langle 3142 \rangle , \, \langle 3214 \rangle, \, \langle 3241 \rangle, \, \langle 3412 \rangle , \,  \langle 3421 \rangle ,  \cr
&{}~{~~~\,~~~} 
\, \langle 4123 \rangle ,  \,   \langle 4132 \rangle, \,   \langle 4213 \rangle, \, \langle 4231 \rangle,  \,    \langle 4312 \rangle, \,    \langle 4321 \rangle ~ \} ~~~,
} \label{eq:S4a}
\ee
or alternately illustrated in the $()$-basis, the twenty-four permutation elements of ${\mathbb{S}}{}_{4}$ are
\be   \eqalign{
\{ {\mathbb{S}}{}_{4} \} ~&=~ \{ ~   (), \, (34),  \,  (23), \, (234) , \,  (243),
\, (24), \, (12), \,  (12)(34),   \,  (123), \,  (1234), \cr
&{}~{~~~~\,~~}  (1243) ,  \,  (124), \,  (132), \,  (1342), \,    (13),   \, (134), \,  (13)(24), \, (1324),  \cr
&{}~{~~~\,~~~}   (1432),   \,   (142), \,  (143), \,  (14) , \,  (1423),  \,  (14)(23)   ~ \} ~~~,
} \label{eq:S4b}
\ee
with the permutation elements being written in the same  order as in Eq.\ (\ref{eq:S4a}).  Together Eq.\ (\ref{eq:S4a})
and Eq.\ (\ref{eq:S4b}) constitute a complete dictionary for translating between expressing the permutation elements 
of ${\mathbb{S}}{}_{4}$ in either the $\langle  \rangle$-notation or the $()$-notation.  A Venn diagram in the form of a circle may 
be used to represent ${\mathbb{S}}{}_{4}$ with the twenty-four permutation elements contained in its interior.
$$
\vCent
{\setlength{\unitlength}{1mm}
\begin{picture}(-20,0)
\put(-44.2,-68.4){\includegraphics[width=3.06in]{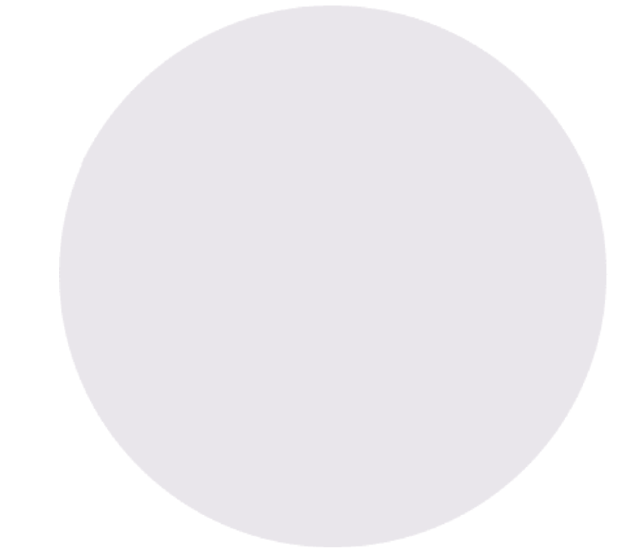}}
\put(-50,-77){{{\bf {Figure}} {\bf {1:}}
A Venn Diagram Representation of ${\mathbb{BC}}{}_{4}$}}
\put(-68.5,-105.3){\includegraphics[width=0.25in]{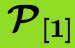}}
\put(-56.2,-105.3){\includegraphics[width=0.25in]{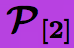}}
\put(-44.1,-105.3){\includegraphics[width=0.25in]{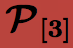}}
\put(-31.9,-105.3){\includegraphics[width=0.25in]{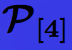}}
\put(-19.6,-105.3){\includegraphics[width=0.25in]{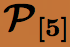}}
\put(1.0,-105.3){\includegraphics[width=0.25in]{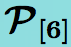}}
\put(-41.7,-113.6){\includegraphics[width=0.25in]{0P1}}
\put(-24.56,-113.6){\includegraphics[width=0.25in]{0P2}}
\put(-7.7,-113.6){\includegraphics[width=0.25in]{0P3}}
\put(9.56,-113.6){\includegraphics[width=0.25in]{0P4}}
\put(26.8,-113.6){\includegraphics[width=0.25in]{0P5}}
\put(44.0,-113.6){\includegraphics[width=0.25in]{0P6}}
\end{picture}}
\nonumber
$$
\vskip3.0in
A fact uncovered by the work in \cite{permutadnk} is that when the 4D, $\cal N$ = 1 supermultiplets are
reduced to 1D, $\cal N$ = 4 supermultiplets, the associated ${\bm {\rL}}{}_{{}_{\rI}}$ and ${\bm {\rR}}{}_{{}_{\rI}}$
matrices require a ``dissection'' of ${\mathbb{S}}{}_{4}$ into six non-overlapping distinct subsets. For reasons
that will be made clear shortly, we denote these subsets by $\{ {~~~~} \}$, $\{ {~~~~} \}$, $\{ {~~~~} \}$, $\{ {~~~~} \}$, 
$\{ {~~~~} \}$, and $\{ {~~~~} \}$ and write the equation
\be{
\{ {\mathbb{S}}{}_{4} \} ~=~ \{ {~~~~} \}
~\cup ~ \{ {~~~~} \}
~\cup ~ \{ {~~~~} \}
~\cup ~ \{ {~~~~} \}
~\cup ~ \{ {~~~~} \}
~\cup ~ \{ {~~~~} \}
~~~,
} \ee
\newpage

where the elements in each subset are given by
$$
\vCent
{\setlength{\unitlength}{1mm}
\begin{picture}(-20,0)
\put(-74,-56){\includegraphics[width=5.6in]{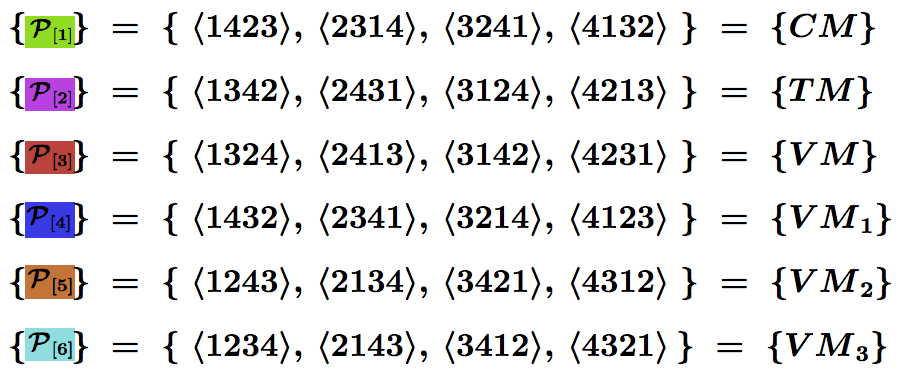}}
\put(-76,-64){{{\bf {Figure}} {\bf {2:}}
A Dissection of ${\mathbb{BC}}{}_{4}$ in Lexicographical Ordering Using $``\langle  \rangle"$ Basis
Elements}}
\end{picture}}
\nonumber
\label{fig:KP}
$$
\vskip2.3in  \noindent
Some explanatory words are in order about this figure.  

The left most column indicates the
distinct subsets.  In the middle of the figure the quartet of permutation elements in each
subset are shown.  In the right most column, the corresponding supermultiplets associated with
the subset are shown.  While the left hand column only contains data associated
with $ \{ {\mathbb{S}}{}_{4} \}$, the right hand column is the result of a specific 
choice of the reduction technique applied to the 4D, $\cal N $ = 1 supermultiplets.
Finally, the $\{ CM \}$, the $\{ TM \}$, and the $\{ VM \}$ labels correspond respectively
to the (S01.), (S05.), and (S07.) supermultiplets described on page two.

After the dissection, the previous Venn diagram can be replaced by one that makes the subsets more obvious.
The color-coding of the notations in Fig.\ 2 indicate
which permutation elements belong to each subset.

$$
\vCent
{\setlength{\unitlength}{1mm}
\begin{picture}(-20,0)
\put(-44,-66){\includegraphics[width=3.2in]{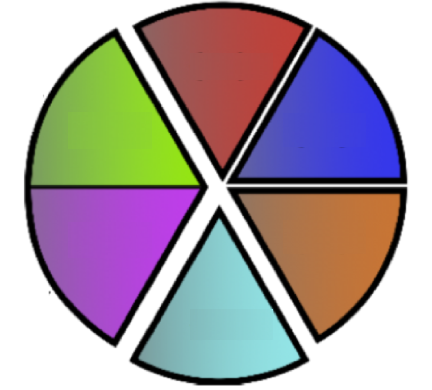}}
\put(-45,-76){{{\bf {Figure}} {\bf {3:}}
``Hodge Duality'' Dissection of ${\mathbb{BC}}{}_{4}$ }}
\end{picture}}
\nonumber
\label{fig:KP}
$$

\newpage

The most obvious feature of the image in Fig.\ 3 is its highly asymmetrical form.  This is due to two 
features. The first can be seen by using the $()$-basis for the permutation elements.  In that convention, 
the various subsets look as shown in Fig.\ 4.

$$
\vCent
{\setlength{\unitlength}{1mm}
\begin{picture}(-20,0)
\put(-54,-64){\includegraphics[width=5.0in]{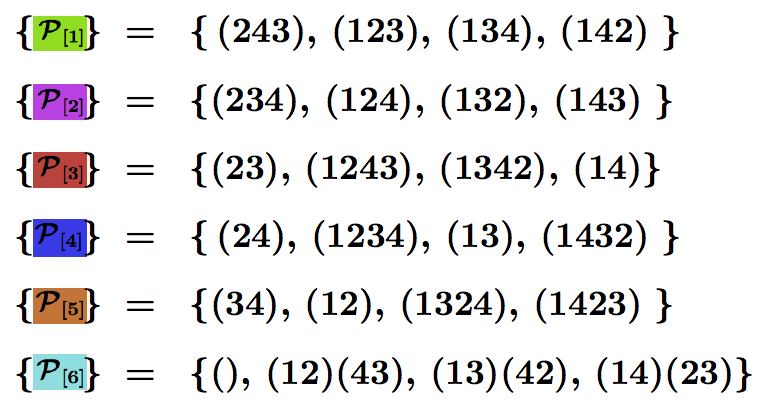}}
\put(-76,-76){{{\bf {Figure}} {\bf {4:}}
A Dissection of ${\mathbb{BC}}{}_{4}$ in Lexicographical Ordering Using $``()"$ Basis Elements}}
\end{picture}}
\nonumber
\label{fig:KP2}
$$
\vskip3.0in
This makes it clear that the first two subsets {\em {only}}
contain permutations that describe 3-cycles.  All the remaining subsets contain either 0-cycles, 2-cycles, 
or 4-cycles.  So the split of the subsets ${\cal P}{}_{[1]}$ and ${\cal P}{}_{[2]}$ away from
${\cal P}{}_{[3]}$, ${\cal P}{}_{[4]}$, ${\cal P}{}_{[5]}$, and ${\cal P}{}_{[6]}$ is actually a split 
between permutations with either even or odd cycles.  None of this is related to the arguments that come 
from SUSY.
 
However, among the permutations that describe even cycles, there is no obvious rationale for
why the groupings above occur. It is the fact that these quartets were derived from projections
of the three supermultiplets (S01.), (S05.), and (S07.) that inspired the arrangements of the 
permutations with even cycles into the groupings shown in $\{ VM \}$, $\{ VM{}_1 \}$, $\{ VM{}_2 \}$,
and $\{ VM{}_3 \}$ 
subsets. Before we continue, it is perhaps useful to recall the arguments related to duality on the SUSY side.

If we begin with a scalar $\Phi$, a quantity $F{}_{\mu}$ can be defined from its derivative. Contracting 
with $F{}^{\mu}$ then leads (with proper normalization) to the Lagrangian proportional to the kinetic 
energy, ${\cal L}{}_{\Phi} $ for a massless scalar field, i.\ e.\
\be
\Phi ~\to ~ \pa{}_{\mu} \Phi ~=~ F{}_{\mu}   ~~~,~~~ {\cal L}{}_{\Phi} ~=~ - \, \fracm 12 \, F{}^{\mu}  \, F{}_{\mu}  ~~~.
\label{eq:phY4}
\ee
There are two differential equations that can be written in terms of $F{}_{\mu}$,
\be
Bianchi ~ Identity ~~~~
\pa{}_{\mu}  F{}_{\nu} ~-~ \pa{}_{\nu} F{}_{\mu} ~=~ 0 ~~~~, ~~~ EoM:~ \pa{}^{\mu} F{}_{\mu} ~=~ 0 ~~~, 
\ee
where the first of these (referred to as the ``$Bianchi \,Identity$'') follows from the definition of $F{}_{\mu}$.  
The second equation (referred to as the ``$EoM$'' or ``$Equation ~of ~Motion$")
arises as the extremum of the field variation applied to ${\cal L}{}_{\Phi}$.

One can repeat line for line the argument above by starting with a 2-form $ B_{{\mu } {\nu} }$
\be
B_{{\mu } {\nu} } ~\to ~ \e{}^{{\mu}{\nu}{\rho}{\sigma} } \pa{}_{\nu}  B_{{\rho } {\sigma} } ~=~ H{}^{\mu}
~~~,~~~
~~~ {\cal L}{}_{B} ~=~  \, \fracm 12 \, H{}^{\mu}  \, H{}_{\mu}  ~~~,
\ee
\be
Bianchi ~ Identity ~~~~
 \pa{}_{\mu}  H{}^{\mu}
 ~=~ 0 ~~~~, ~~~ EoM:~ \pa{}_{\mu}  H{}_{\nu} ~-~ \pa{}_{\nu} H{}_{\mu} ~=~ 0
~~~. \ee
It is thus clear that the final difference in starting from $\Phi$ versus starting from $ B_{{\mu } {\nu} }$ is
the exchange of the ``$Bianchi \,Identity$'' and
the ``$EoM$.''

The behavior of the 0-form $\Phi$ and the 2-form $B{}_{\m \n} $ contrast greatly with the behavior of the 1-form $A{}_{\m}$.  Following the same reasoning as above, one is led to
\be
A{}_{\mu} ~\to ~ \pa{}_{\mu} A{}_{\nu} ~-~ \pa{}_{\nu} A{}_{\mu}
~=~ F{}_{{\mu} {\nu}}    ~~~,~~~ {\cal L}{}_{\Phi} ~=~ - \, \fracm 14 \, F{}^{{\mu}{\nu}}  \, F{}_{{\mu} {\nu}}
\ee
and 
\be
Bianchi ~ Identity ~~~~
\e{}^{{\mu} {\nu} {\rho} {\sigma}}
\pa{}_{\mu}  F{}_{{\nu}{\rho}}   ~=~ 0 ~~~~, ~~~ EoM:~ \pa{}^{\mu} F{}_{{\mu} {\nu}} ~=~ 0
\label{eq:dphtN}
\ee
now we treat the $EoM$ as a constraint to see if it has a solution.  It is easily shown that 
\be{
F{}_{\m \n} ~\equiv~ \e{}_{\m \n}{}^{\rho \s}
\pa{}_{\rho} b{}_{\sigma}
} \label{eq:phY10}
\ee
solves the second equation in (\ref{eq:dphtN}) and upon substituting this into the $Bianchi~ Identity$, one finds $\pa{}^{\m} [\, \pa{}_{\mu} b{}_{\nu} ~-~ \pa{}_{\nu} b{}_{\mu} \,]  $ = 0.  So the exchange of the $EoM$ and the $Bianchi ~Identity $ occurs, but the dual field $ b{}_{\mu}$ is also a 1-form.

Motivated by these arguments, the work in \cite{permutadnk} contained a proposal that there should be a
mapping operation acting on the ${\mathbb{S}}{}_{4}$-space of permutations that is the ``shadow" of the
behavior discussed in Eq.\ (\ref{eq:phY4}) - Eq.\ (\ref{eq:phY10}).  It turns out that the realization
of such a mapping operation (denoted by ${}^*$) is straightforward.  The $()$-notation shown in Fig.\ 4
is most efficient in this realization.

The ${}^*$ map is defined by simply reversing the direction of reading the cycles in $()$-notation for the
permutations.  Thus, we define
\begin{equation}
    \begin{split}
         {\bm {  {~}^{*} \{ }\rm  CM} {\bm {\}} }~~\bm \equiv&~  {\bm {\{  \, (234)~~~~ , \,  ~(213)~~~, \, (143)~~\, , \, (124)    \,  \} }} \\
         {\bm {  {~}^{*} \{ }\rm TM} {\bm {\}} }~~\bm \equiv&~  {\bm {\{  \, (243) ~~~\,\, , \,  ~(142)~\,~\,, \, (123)~~\,, \, (134)    \,  \} }} \\
         {\bm {  {~}^{*} \{ }\rm VM} {\bm {\}} }~~\bm \equiv&~  {\bm {\{  \, (1342)~~\,  , \,  ~~(23)~~\,~, \, ~(14)~~~, \, (1243)    \,  \} }} \\
         {\bm {   {~}^{*} \{ }\rm VM{}_1} {\bm {\}} }~~\bm \equiv&~  {\bm {\{  \, (1234)~~\,  , \,  ~~(24) ~~\,~\, , \, (1432)\, , \, (13)    \,  \} }}\\
          {\bm {  {~}^{*} \{ }\rm VM{}_2} {\bm {\}} }~~\bm \equiv&~   {\bm {\{  \, (1423)~~\, , \,  ~(1324)~\, , \, ~~(12)~~, \, (34)    \,  \} }} \\
          {\bm {  {~}^{*} \{ }\rm VM{}_3} {\bm {\}} }~~\bm \equiv&~  {\bm {\{  \, (13)(24) , \,  (14)(23), \, ~~~()~~~~ , \, (12)(34)    \,  \} }}
    \end{split}
\end{equation}

and if we consider only unordered sets these imply
\be {
{\bm {  {~}^{*} \{ } CM} {\bm {\}} } ~=~ {\bm {   \{ } TM} {\bm {\}} } ~~,~~ 
{\bm {  {~}^{*} \{ } TM} {\bm {\}} } ~=~ {\bm {   \{ } CM} {\bm {\}} } ~~,~~ 
{\bm {  {~}^{*} \{ } VM} {\bm {\}} } ~=~ {\bm {   \{ } VM} {\bm {\}} } ~~,~~
~~,~~
} \ee
for the permutations associated with the (S01.), (S05.) and (S07.) supermultiplets. Furthermore
\be{
{\bm {  {~}^{*} \{ } VM{}_1} {\bm {\}} } ~=~ {\bm {   \{ } VM{}_1} {\bm {\}} }
~~,~~
{\bm {  {~}^{*} \{ } VM{}_2} {\bm {\}} } ~=~ {\bm {   \{ } VM{}_2} {\bm {\}} }
~~,~~
{\bm {  {~}^{*} \{ } VM{}_3} {\bm {\}} } ~=~ {\bm {  \{ } VM{}_3} {\bm {\}} }
~~.~~
}
\ee
for the permutations that were not constructed from projections.  These results can be summarized
in the Venn diagram shown in Fig.\ 5
\newpage
$$
\vCent
{\setlength{\unitlength}{1mm}
\begin{picture}(-20,0)
\put(-65,-93){\includegraphics[width=4.3in]{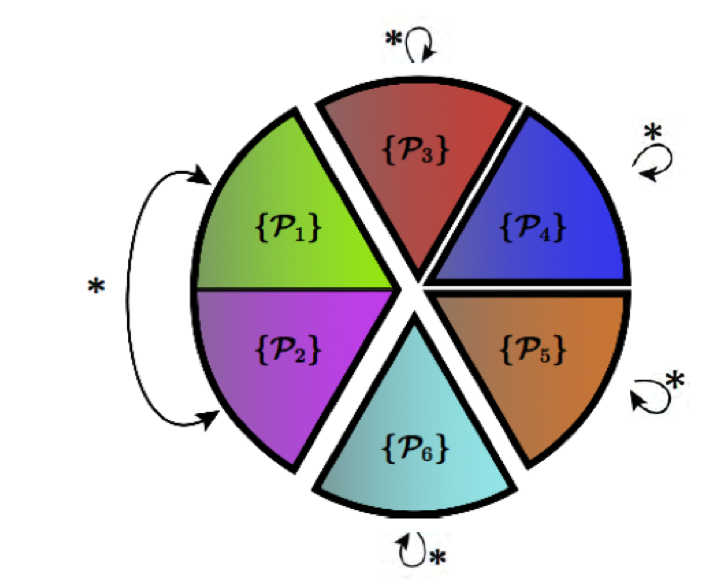}}
\put(-45,-103){{{\bf {Figure}} {\bf {5:}}
``Hodge Duality'' Mapping of ${\mathbb{BC}}{}_{4}$ }}
\end{picture}}
\nonumber
\label{fig:KPp}
$$
\vskip4.0in

It is important to realize that the definition of the ${}^*$ operation acting on
${\bm {\rL}} {}_{{}_{\rI}}$ matrices is {\em {not}} unique.  In principle, one can define the
${}^*$ operation by first performing a ${{\mathbb{O}}}(4) \times
{{\mathbb{O}}}(4)$ transformation 
\be \eqalign{
{\bm {\rL}} {}_{{}_{\rI}} ~\to~ {\cal  X} \, {\bm {\rL}} {}_{{}_{\rI}}   \, {\cal Y}
} ~~~,
\label{eq:E03}
\ee
where
\be \eqalign{
{\cal  X} \, ({\cal X})^t ~=~   ({\cal X})^t  \, {\cal  X} ~=~
{\cal  Y} \, ({\cal Y})^t ~=~   ({\cal Y})^t  \, {\cal  Y} 
~=~    {\bm {\mathbb{I}}}{}_4  ~~~,
}
\label{eq:E04}
\ee
and only afterward then read the $()$-basis cycles in reverse order.  The definition of
the ${}^*$ operation we use is simply the minimal one, but more baroque ones are also
consistent.

However, though the subsets may now be consistently interpreted as projections of the
supermultiplets, there still remains a mystery.  If one does not invoke the argument
about projection from supermultiplets, what structure in ${\mathbb{S}}{}_{4}$ is 
operative without an appeal to SUSY to determine these quartets?

In the remainder of this work, we will propose an answer to this question.

\newpage
\section{Bruhat Weak Ordering and the Permutahedron}
\label{sec:DeFNs}

\subsection{Bruhat Weak Ordering}
\label{sec:BrUHT}

The concept of the {\em {Hamming distance}} between two strings of bits \cite{HaM} is a
well known one for bitstrings of equal length.

One simply counts the number of bits that
need to be ``flipped'' to go from one string to another.  In this sense, one can regard
the Hamming distance as a choice of a metric on the space of bitstrings.

Within the realm of the mathematics of permutation groups, there exist the concepts of
{\em {weak left Bruhat ordering}} and {\em {weak right Bruhat ordering}}.  The mathematician Adaya-Nand Verma (1933-2012) initiated the combinatorial study of 
Bruhat order and named it after earlier work completed by mathematician Francois Bruhat (1929-2017). Further details about weak left- and right- Bruhat ordering can be found in the work of the work of \cite{BruHT}.

The realization of either weak
left- or right- Bruhat ordering can be used to define a metric on the space of elements of a permutation group. In this ``spirit,'' at least for our purposes, the existence of these ordering prescriptions for the elements of the permutation group, is rather
similar to the role of the Hamming distance 
on the space of bitstrings. We will begin by
illustrating our use of weak ordering in
Fig.\ 6 and then turn to an explanation of the image.
$$
\vCent
{\setlength{\unitlength}{1mm}
\begin{picture}(-20,0)
\put(-52,-106){\includegraphics[width=4.6in]{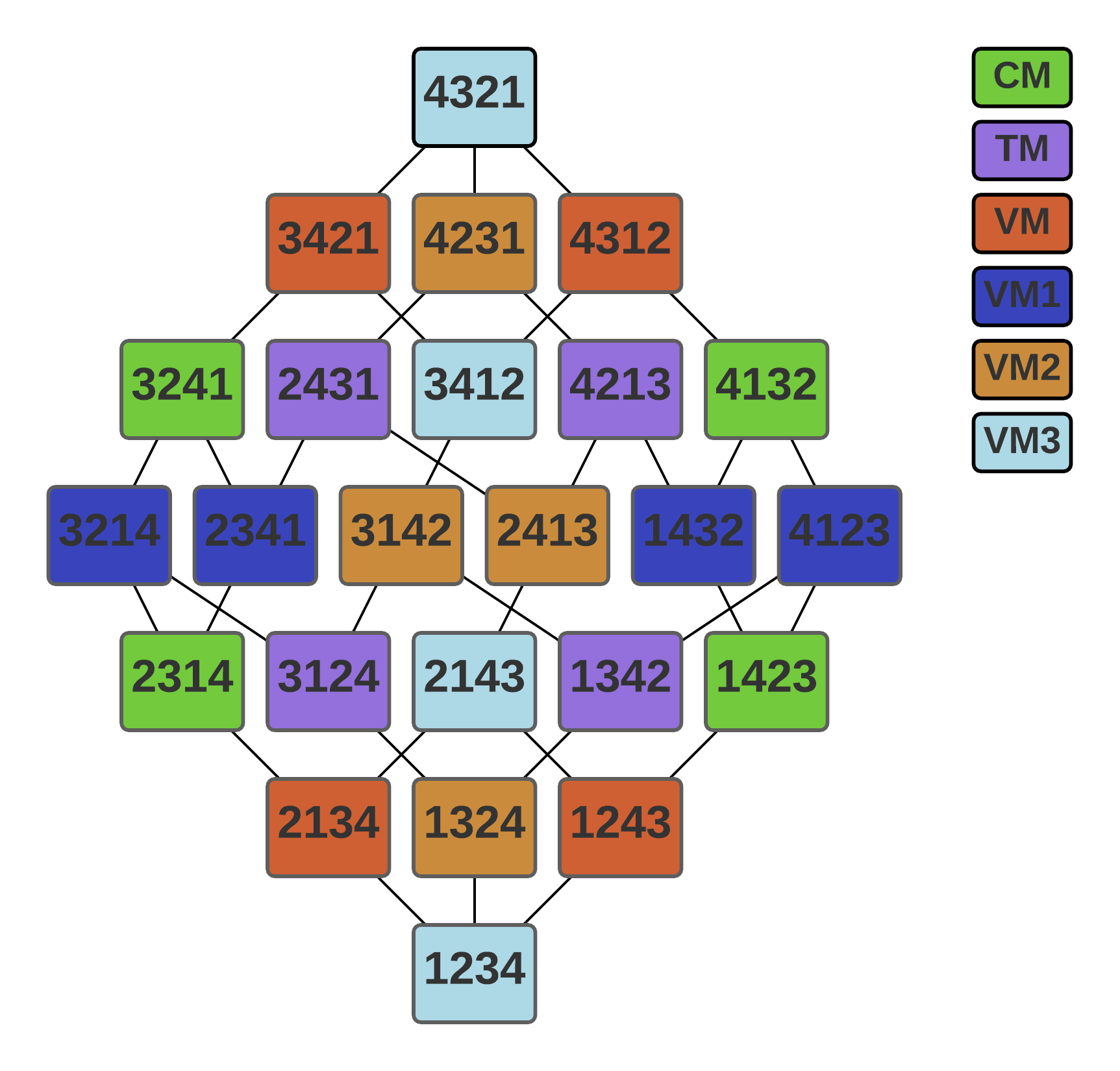}}
\put(-32,-111){{{\bf {Figure}} {\bf {6:}}
Weak Ordering in ${\mathbb{BC}}{}_{4}$ }}
\end{picture}}
\nonumber
\label{fig:OP}
$$
\vskip4.2in
\noindent
On the right most portion of the image, illustrated in a color-coded column, are the six subsets containing the
quartets.  Distributed throughout the
main portion of the figure are the individual permutation elements listed by
their ``bra-ket'' addresses and enclosed by
the color of the quartet subset to which each is a member.  Once this figure is obtained it immediately implies an intrinsic definition of a ``minimal Bruhat distance''
between any two elements by simply counting the number of links that connect the two elements along a minimum path.

There remains the task of explaining how this ordering was obtained.

To describe the procedure, let us start with the identity permutation that is illustrated
as the bottom most permutation in the figure.

The identity permutation element has the ``bra-ket address"
$\langle 1234 \rangle$.  There are three pair-wise adjacent 2-cycle permutations $(12)$, $(23)$, and $(34)$ that respectively
provide instructions to change the first two numbers in the address, the second two numbers in the
address, and finally the last two numbers in the address.  The $(12)$ permutation will send the identity
address into the left most address at the first elevated level of Fig.\ 6.  The $(23)$ permutation will send the 
identity address into the middle address at the first elevated level of Fig.\ 6. Finally, the $(34)$ permutation 
will send the identity address into the right most address at the first elevated level of Fig.\ 6.  Next the
three permutations $(12)$, $(23)$, and $(34)$ are applied to the three addresses at the first elevated
level.  Sometimes, this will result in a restoration of the identity address.  However, when this does
not occur, the new addresses will be those that occur at the second elevated level.  One simply repeats
this procedure until the addresses of all the permutations have been reached.  Thus, the links shown in
the network of Fig.\ 6. represent the orbits of the addresses of the permutation elements under the action 
of the three pair-wise adjacent 2-cycle permutations $(12)$, $(23)$, and $(34)$.

\subsection{The Permutahedron}
\label{sec:pMUT}

In the mathematics literature, the concept of permutahedra seems
first to have been a topic of study carried out by Pieter Schoute
(1846 -1923) in 1911 \cite{pHR0n1}.  Further informative readings on this subject can be found in the references \cite{pHR0n2,pHR0n3,pHR0n4,pHR0n5}.
$$
\vCent
{\setlength{\unitlength}{1mm}
\begin{picture}(-20,0)
\put(-41.0,-64.0){\includegraphics[width=2.5in]{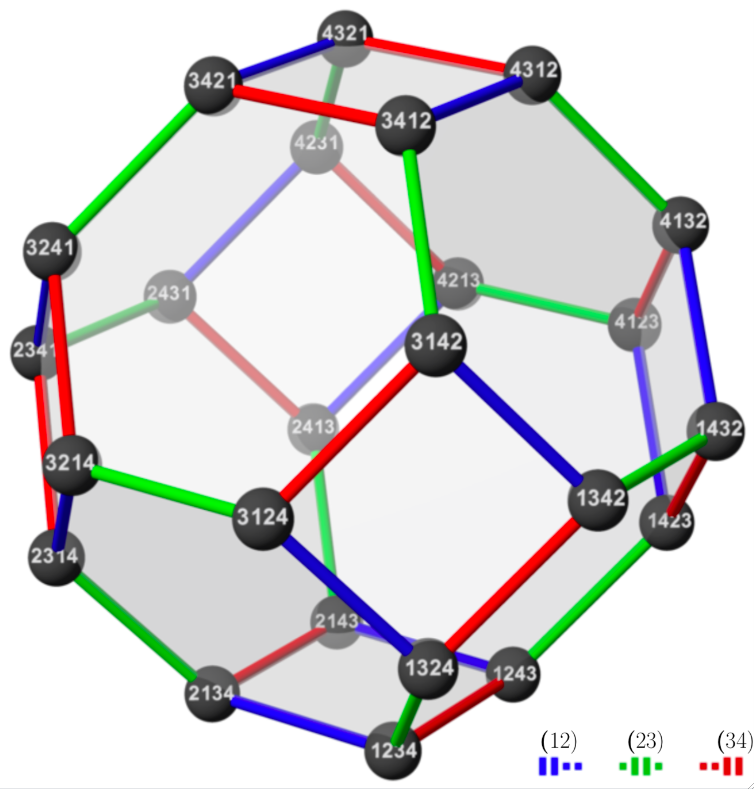}}
\put(3,-65){\includegraphics[width=2.0in]{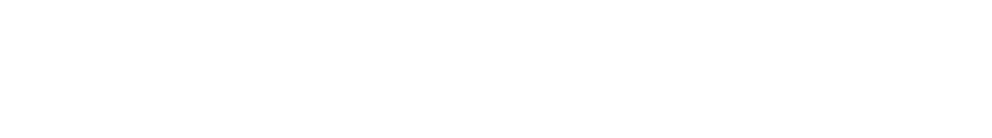}}
\put(12,-65){\includegraphics[width=1.4in]{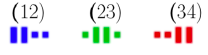}}
\put(-30,-70.4){\includegraphics[width=2.0in]{WHTmm}}
\put(-60,-70.4){\includegraphics[width=2.0in]{WHTmm}}
\put(-70.0,-74.0){{{\bf {Figure}} {\bf {7:}}
${\mathbb{S}}{}_{4}$ Permutation Elements and the Truncated Octahedron}}
\end{picture}}
\nonumber
$$
\vskip2.8in
The appropriate permutahedron, in the context we require is a polytope that can be constructed
by using the network shown in Fig.\ 6 and conceiving of that image as a net.  One can begin with a truncated
octahedron\footnote{A simplified discussion of the truncated octahedron can be found at $~~~~~~~~~~~~~~~~~~~~~~~~~~~~~~$
https://en.wikipedia.org/wiki/Truncated$\_$octahedron} where each  vertex
of the truncated octahedron represents a permutation element.

The colors of the links are related to the three permutations links (12), (23), and (34) respective by blue, green, and red as shown by the epigram to the lower right side of Fig.\ 7.

The network shown in Fig.\ 6 perfectly aligns with the permutahedron shown in Fig.\ 7.

A way to see this is to begin with the image in Fig.\ 7 and treat it as a frame that can be `adorned' by the individual permutations listed in Fig.\ 2. By matching each ``bra-ket address,''
one is led to a figurative ``Christmas tree'' seen in Fig.\ 8.

$$
\vCent
{\setlength{\unitlength}{1mm}
\begin{picture}(-20,0)
\put(-57,-95){\includegraphics[width=4.7in]{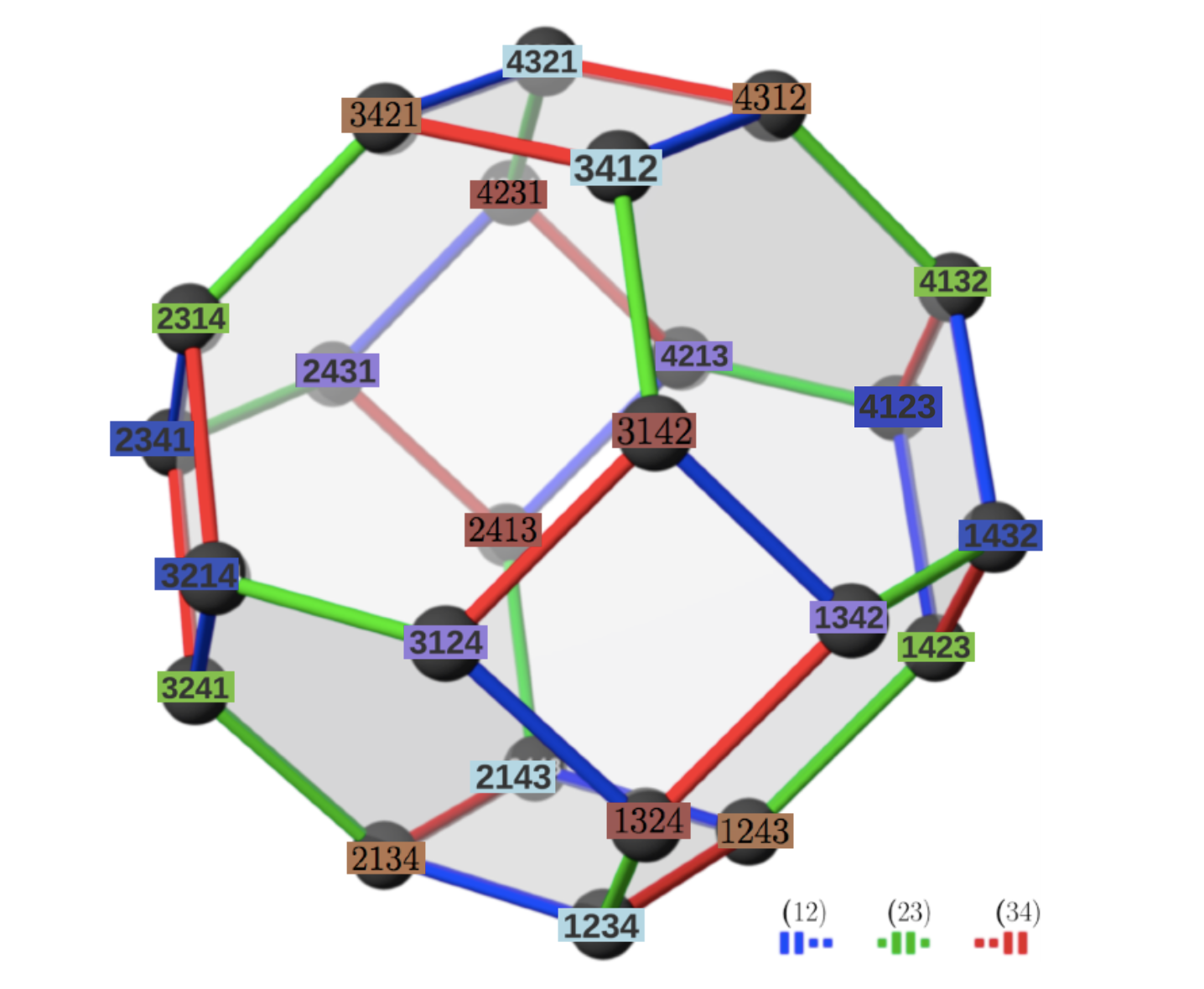}}
\put(-70,-105){{{\bf {Figure}} {\bf {8:}  ${\bm {\{ {\cal P}}{}_{[1]} \}}$ }} $\cdots$ 
${\bm {\{ {\cal P}}{}_{[6]} \}}$ addresses adorning the Permutahedron }
\put(-47.2,-106.4){\includegraphics[width=0.3in]{0p1}}
\put(-25.4,-106.5){\includegraphics[width=0.3in]{0p6}}
\end{picture}}
\nonumber
\label{fig:Xmas}
$$
\vskip4.2in
\noindent
Having completely adorned the permutahedron vertices with the color-coded labels of the individual permutations
from the subsets, one can check that the thirty-six edges connect the vertices in precisely the same
manner as shown in the network of Fig.\ 6.

From a visual examination of Fig.\ 8, it is clear that the subset elements ${\bm\{}$\raisebox{-2mm}{\includegraphics[width=0.3in]{0p1}}${\bm\}}$
$\cdots$ ${\bm\{}$\raisebox{-2mm}{\includegraphics[width=0.3in]{0p6}}${\bm\}}$
occur in a highly symmetrical arrangement of the
four colors about the adorned permutahedron.  So the next obvious step is to mathematically quantify
this symmetrical arrangement of the four colors.

Stated in a different fashion, the appearance of the 4D, $\cal N$ = 1 supersymmetrical minimal quartets is 
equivalent to solving a coloring problem.  Start with four colors and the truncated octahedron {\em {with 
no labeling of nodes}}, by using the colors, what is the maximally symmetrical choice for painting the 
nodes?  The answer is the SUSY quartets.

\newpage
\subsection{ The Permutahedron, $\cal X$, and $\cal Y$ Transformations}
\label{sec:CalcCaly}

The existence of the permutahedron also brings into clear focus the meaning of the
transformations shown in Eq.\ (\ref{eq:E03}) and Eq.\ (\ref{eq:E04}).  Let us illustrate
this in the case of one example. Consider only the ${\bm {\cal P}}{}_{[6]} $ (illustrated in Fig.\ 14 below).  If the $(12)$ permutation
is applied to it, the four light blue nodes shown in  Fig.\ 14 ``move'' to the locations of the brown
nodes in Fig.\ 13.  If the  $(23)$ permutation is applied after the $(12)$ permutation, then the four light blue nodes shown in  Fig.\ 5 ``move'' to the locations of the brown nodes in Fig.\ 9. The $\cal X$, and $\cal Y$ transformation correspond to left or right multiplication of the
elements of the quartets by different orders and power of the permutation elements
$(12)$, $(23)$, or $(34)$.

The remainder of this paper is devoted to the mathematical calculation that quantifies all these observations
of this chapter.

\newpage
\section{The 300 ``Correlators'' of the $\bm {{\mathbb{S}}{}_{4}}$ Permutahedron}
\label{sec:DeFNP1}

In the previous chapters, the presentation contained arguments and visual images based only on the properties 
of the truncated octahedron to propose that the $ {\bm {\rL}}{}_{{}_{\rI}}$ matrices that can be
used to define SUSY systems in 1D, are in fact solutions to a four color problem on the truncated octahedron.
From here on, we will present the calculational evidence that supports the discussions in the earlier chapters.

For this purpose, there is introduced a function denoted by ${\bm {\cal A}}{}_{\ell x}[{\cal P}{}_{[a|A]},\, {\cal P}{}_{[b|B]} ]$ 
that we will call a ``two-point correlator.''  This function assigns a number of the minimal
Bruhat distance between the elements ${\cal P}{}_{[a|A]}$  and $ {\cal P}{}_{[b|B]} $ contained in 
$\bm {{\mathbb{S}}{}_{4}}$. This means this is a symmetric 24 $\times$ 24 matrix with 300 possible
entries.  The symbol ${\cal P}{}_{[a|A]}$ is meant to denote the $a$-th element
in the dissected subset $A$.  If the two subsets are such that $A$ = $B$, we will call these ``intra-quartet
correlators", and if the two subsets are such that $A \ne B$, we then call these ``inter-quartet
correlators."

We can begin our calculations by considering the case where $A$ = $B$ = 1 that corresponds to the 
first of the dissected subsets according to Fig.\ 2.  The four permutation elements belonging to 
this subset have the corresponding vertices ``blocked out'' by green quadrilaterals in Fig. \ref{fig:P-cm}.


$$
\vCent
{\setlength{\unitlength}{1mm}
\begin{picture}(-20,0)
\put(-60,-129){\includegraphics[width=4.6in]{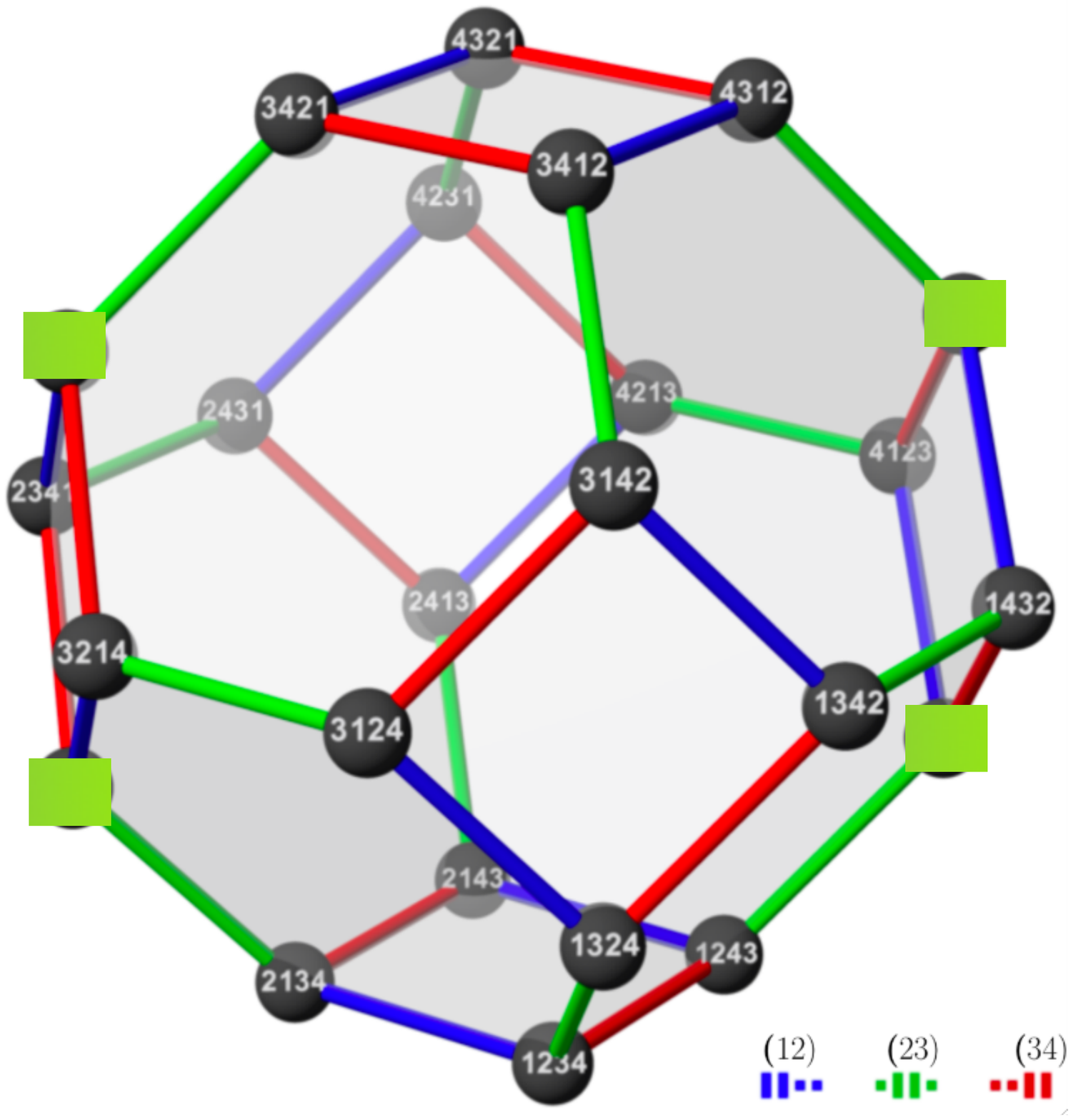}}
\put(-50,-120){\includegraphics[width=4.6in]{WHTmm}}
\put(-50,-114){{{\bf {Figure}} {\bf {9:}}
Permutahedron \& Colored ${\cal P}{}_{[1]}$ Subset Elements }}
\end{picture}}
\nonumber
\label{fig:P-cm}
$$
\vskip4.5in

From simply counting the number of links it takes to ``travel'' 
from a given specified element $a$ in the set ${\cal P}{}_{[1]}$ 
to a second specified element $b$ in ${\cal P}{}_{[1]}$ we are 
able to create a table.  The rows of the table are listed according to
the lexicographical ordering of the element in ${\cal P}{}_{[1]}$ 
and the columns of the table are listed according to the lexicographical 
ordering of the elements in ${\cal P}{}_{[1]}$ also.

\begin{equation}
\begin{tabular}{|c|ccccccccccc|} \hline
\diagbox{$\bm {{\cal P}{}_{[1]}}$}{$\bm {{\cal P}{}_{[1]}} $} &{~}& \,  
$\langle 1423 \rangle$  & {~} & {~} $\langle 2314 \rangle$ & {~} & {~} 
$\langle 3241 \rangle$  &{~}& &$\langle 4132 \rangle$ & & \\ \hline
$\langle 1423 \rangle$  &{~}& \,  0 & {~} & {~}4 & {~} & {~}6 &{~}& &2& & \\ \hline
$\langle 2314 \rangle$  &{~}& \,  4 & {~} & {~}0 & {~} & {~}2 &{~}& &6& & \\ \hline
$\langle 3241 \rangle$  &{~}& \,  6 & {~} & {~}2 & {~} & {~}0 &{~}& &4& & \\ \hline
$\langle 4132 \rangle$  &{~}& \,  2 & {~} & {~}6 & {~} & {~}4 &{~}& &0& & \\ \hline
\end{tabular}
\label{N81}
\end{equation}
\begin{center}
{{\bf Table 1:} $\{CM\}-\{CM\}~ {\rm {Two-Point ~  Correlator~ Values}}$}
 \end{center}

The information in the table above can be used to determine the values of 
${\bm {\cal A}}{}_{\ell x}[{\cal P}{}_{[a|1]},\, {\cal P}{}_{[b|1]} ] $ in 
the form of a matrix.
\be
{\bm {\cal A}}{}_{\ell x}[{\cal P}{}_{[a|1]},\, {\cal P}{}_{[b|1]} ] ~=~
\left[\begin{array}{cccc}
~0 & ~~4 &  ~~6  &  ~~2\\
~4 & ~~0 &  ~~2  &  ~~6\\
~6 & ~~2 &  ~~0  &  ~~4\\
~2 & ~~6 &  ~~4  &  ~~0\\
\end{array}\right]  {~~~~~~~~~~~}  {~~~~~~~~~~}
\label{eq:Mtrx1}
\ee

It is clear that the trace of this matrix vanishes and a direct calculation of its eigenvalues
yield $\{ 12, \, 0, \,  -4 , \, -8
\}$ expressed in descending order.

There are five more cases  ((\ref{eq:Mtrx2}), (\ref{eq:Mtrx3}), (\ref{eq:Mtrx4}), (\ref{eq:Mtrx5}), and (\ref{eq:Mtrx6})) where $A$ = $B$.  Direct calculations for each of these yield the
same value for the traces as well for the sets of eigenvalues.  This is true even though the six matrices
involved are all distinct.  However, the six matrices possess many similarities.  For instance each one
contains six entries that are either 2's, 4's or 6's.  In no row or column do these numbers appear twice.  There are exactly 3! = 6 ways to satisfy this condition, hence six quartets.  

Intuitively, something like this should have been an expectation by looking back at the image in 
Fig.\ 8.  If we assign the standard right-handed $x$-$y$-$z$ coordinate axes to the image in Fig.\ 8,
where $x$-axis points out of the paper, it is apparent that each quartet has two of its members on the opposite faces of square faces that are perpendicular to the coordinate axes.

In the remainder of this subsection, we will illustrate the explicit analogous results for the remaining
cases were $A$ = $B$.
\newpage

\newpage
\subsection{Intra-quartet  $\bm {{\cal P} {}_{[2]}} $ Correlators of the $\bm {{\mathbb{S}}{}_{4}}$ Permutahedron}
\label{sec:DeFNP2}

In this subsection, we illustrate the intra-quartet 
two-point correlators associated with
$\bm {{\cal P} {}_{[2]}} $.  The quartet members are indicated by purple quadrilaterals on some of the
vertices of the permutahedron, the correlators are shown in tabular form, and finally this same data is presented as a matrix.

$$
\vCent
{\setlength{\unitlength}{1mm}
\begin{picture}(-20,0)
\put(-60,-120){\includegraphics[width=4.6in]{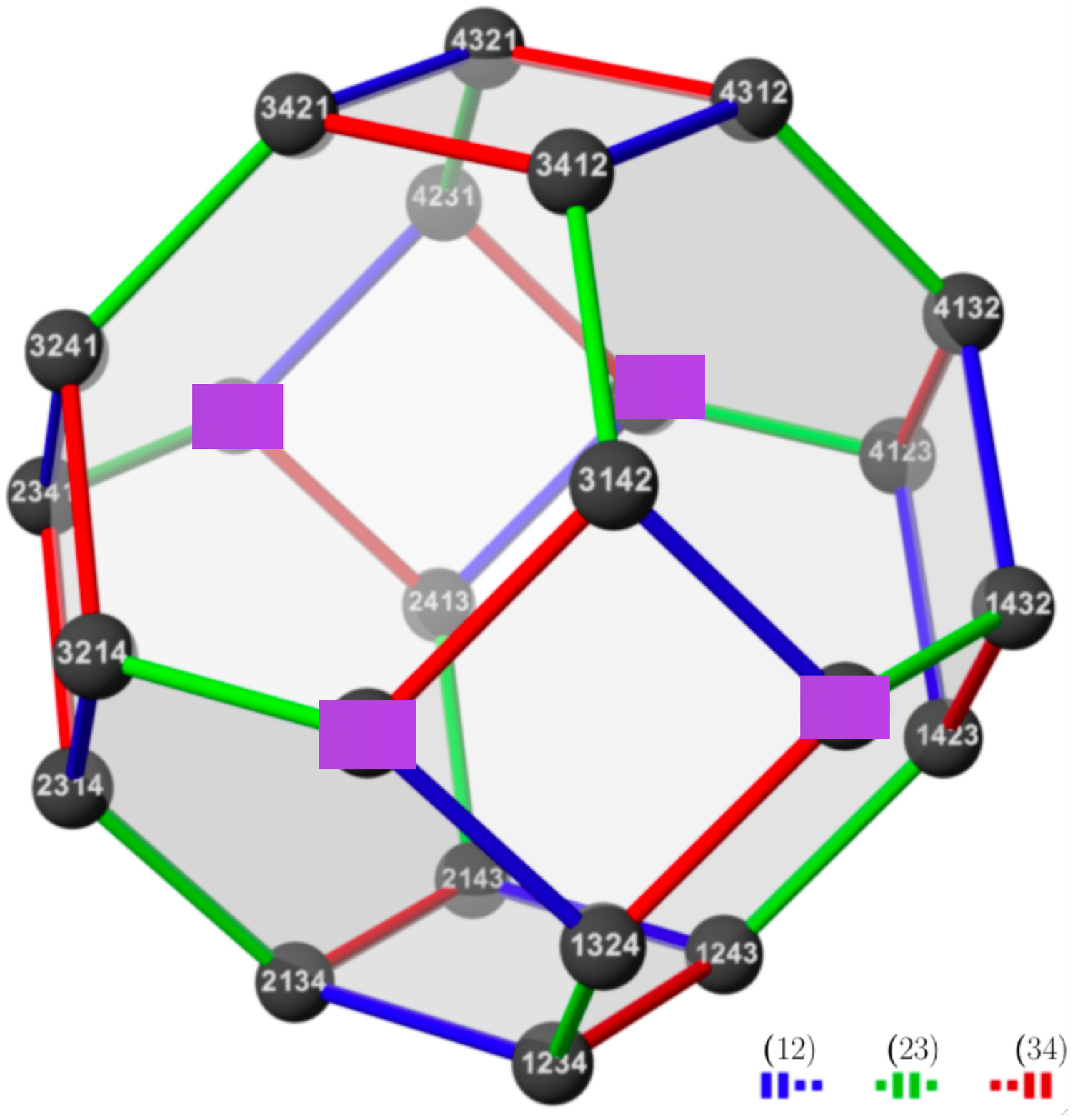}}
\put(-50,-104){{{\bf {Figure}} {\bf {10:}}
Permutahedron \& $ {\cal P}{}_{[2]}  $ Subset Elements }}
\end{picture}}
\nonumber
\label{fig:P-tm}
$$
\vskip4.2in
The data contained in 
${\bm {\cal A}}{}_{\ell x}[{\cal P}{}_{[a|2]},\, {\cal P}{}_{[b|2]} ] $ is presented below first in tabular 
\begin{equation}
\begin{tabular}{|c|ccccccccccc|} \hline
\diagbox{$\bm {{\cal P}{}_{[2]}}$}{$\bm {{\cal P}{}_{[2]}} $} &{~}& \,
$\langle 1342 \rangle$ & {~} & {~}$\langle 2431 \rangle$ & {~} & {~}$\langle 3124 \rangle$ &{~}& &$\langle 4213 \rangle$& & \\ \hline
$\langle 1342 \rangle$ &{~}& \,  0 & {~} & {~}6 & {~} & {~}2 &{~}& &4& & \\ \hline
$\langle 2431 \rangle$ &{~}& \,  6 & {~} & {~}0 & {~} & {~}4 &{~}& &2& & \\ \hline
$\langle 3124 \rangle$ &{~}& \,  2 & {~} & {~}4 & {~} & {~}0 &{~}& &6& & \\ \hline
$\langle 4213 \rangle$ &{~}& \,  4 & {~} & {~}2 & {~} & {~}6 &{~}& &0& & \\ \hline
\end{tabular}
\label{282}
\end{equation}  
\begin{center}
{{\bf Table 2:} $\{TM\}-\{TM\}~ {\rm {Two-Point ~  Correlator~ Values}}$}
 \end{center}
and in matrix form.
\be
{\bm {\cal A}}{}_{\ell x}[{\cal P}{}_{[a|2]},\, {\cal P}{}_{[b|2]} ] ~=~
\left[\begin{array}{cccc}
~0 & ~~6 &  ~~2  &  ~~4\\
~6 & ~~0 &  ~~4  &  ~~2\\
~2 & ~~4 &  ~~0  &  ~~6\\
~4 & ~~2 &  ~~6  &  ~~0
\end{array}\right]  {~~~~~~~~~~~}  {~~~~~~~~~~}
\label{eq:Mtrx2}
\ee

\newpage
\subsection{Intra-quartet  $\bm {{\cal P} {}_{[3]}} $ Correlators of the $\bm {{\mathbb{S}}{}_{4}}$ Permutahedron}
\label{sec:DeFNP3}

In this subsection, we illustrate the intra-quartet two-point correlators associated with $\bm {{\cal P} {}_{[3]}} $.  
The quartet members are indicated by rust color quadrilaterals on some of the vertices of the permutahedron, the 
correlators are shown in tabular form, and finally this same data is presented as a matrix.

$$
\vCent
{\setlength{\unitlength}{1mm}
\begin{picture}(-20,0)
\put(-60,-120){\includegraphics[width=4.6in]{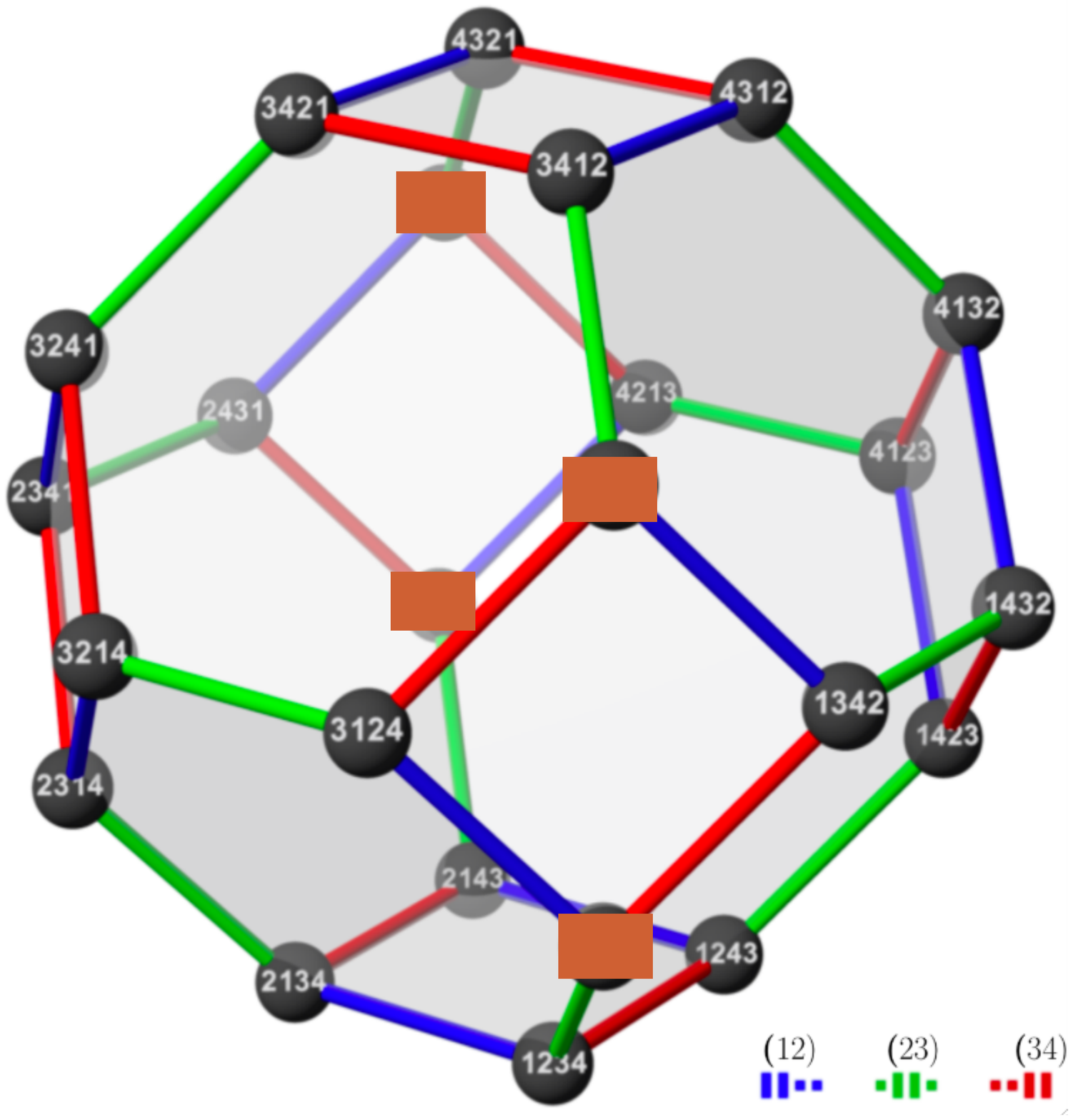}}
\put(-50,-104){{{\bf {Figure}} {\bf {11:}}
Permutahedron \& $ {\cal P}{}_{[3]}  $ Subset Elements }}
\end{picture}}
\nonumber
\label{fig:P-vm}
$$
\vskip4.2in
The data contained in
${\bm {\cal A}}{}_{\ell x}[{\cal P}{}_{[a|3]},\, {\cal P}{}_{[b|3]} ] $ is presented below first in tabular 
\begin{equation}
\begin{tabular}{|c|ccccccccccc|} \hline
\diagbox{$\bm {{\cal P}{}_{[3]}}$}{$\bm {{\cal P}{}_{[3]}} $} &{~}& \,
$\langle 1324 \rangle$ & {~} & {~}$\langle 2413 \rangle$ & {~} & {~}$\langle 3142 \rangle$ &{~}& &$\langle 4231 \rangle$& & \\ \hline
$\langle 1324 \rangle$ &{~}& \,  0 & {~} & {~}4 & {~} & {~}2 &{~}& &6& & \\ \hline
$\langle 2413 \rangle$ &{~}& \,  4 & {~} & {~}0 & {~} & {~}6 &{~}& &2& & \\ \hline
$\langle 3142 \rangle$ &{~}& \,  2 & {~} & {~}6 & {~} & {~}0 &{~}& &4& & \\ \hline
$\langle 4231 \rangle$ &{~}& \,  6 & {~} & {~}2 & {~} & {~}4 &{~}& &0& & \\ \hline
\end{tabular}
\label{282}
\end{equation}  
\begin{center}
{{\bf Table 3:} $\{VM\}-\{VM\}~ {\rm {Two-Point ~  Correlator~ Values}}$}
 \end{center}
and in matrix form.
\be
{\bm {\cal A}}{}_{\ell x}[{\cal P}{}_{[a|3]},\, {\cal P}{}_{[b|3]} ] ~=~
\left[\begin{array}{cccc}
~0 & ~~4 &  ~~2  &  ~~6\\
~4 & ~~0 &  ~~6  &  ~~2\\
~2 & ~~6 &  ~~0  &  ~~4\\
~6 & ~~2 &  ~~4  &  ~~0
\end{array}\right]  {~~~~~~~~~~~}  {~~~~~~~~~~}
\label{eq:Mtrx3}
\ee

\newpage
\subsection{Intra-quartet  $\bm {{\cal P} {}_{[4]}} $ Correlators of the $\bm {{\mathbb{S}}{}_{4}}$ Permutahedron}
\label{sec:DeFNP4}

In this subsection, we illustrate the intra-quartet two-point correlators associated with $\bm {{\cal P} {}_{[4]}} $.  
The quartet members are indicated by dark blue quadrilaterals on some of the vertices of the permutahedron, the 
correlators are shown in tabular form, and finally this same data is presented as a matrix.

$$
\vCent
{\setlength{\unitlength}{1mm}
\begin{picture}(-20,0)
\put(-60,-120){\includegraphics[width=4.6in]{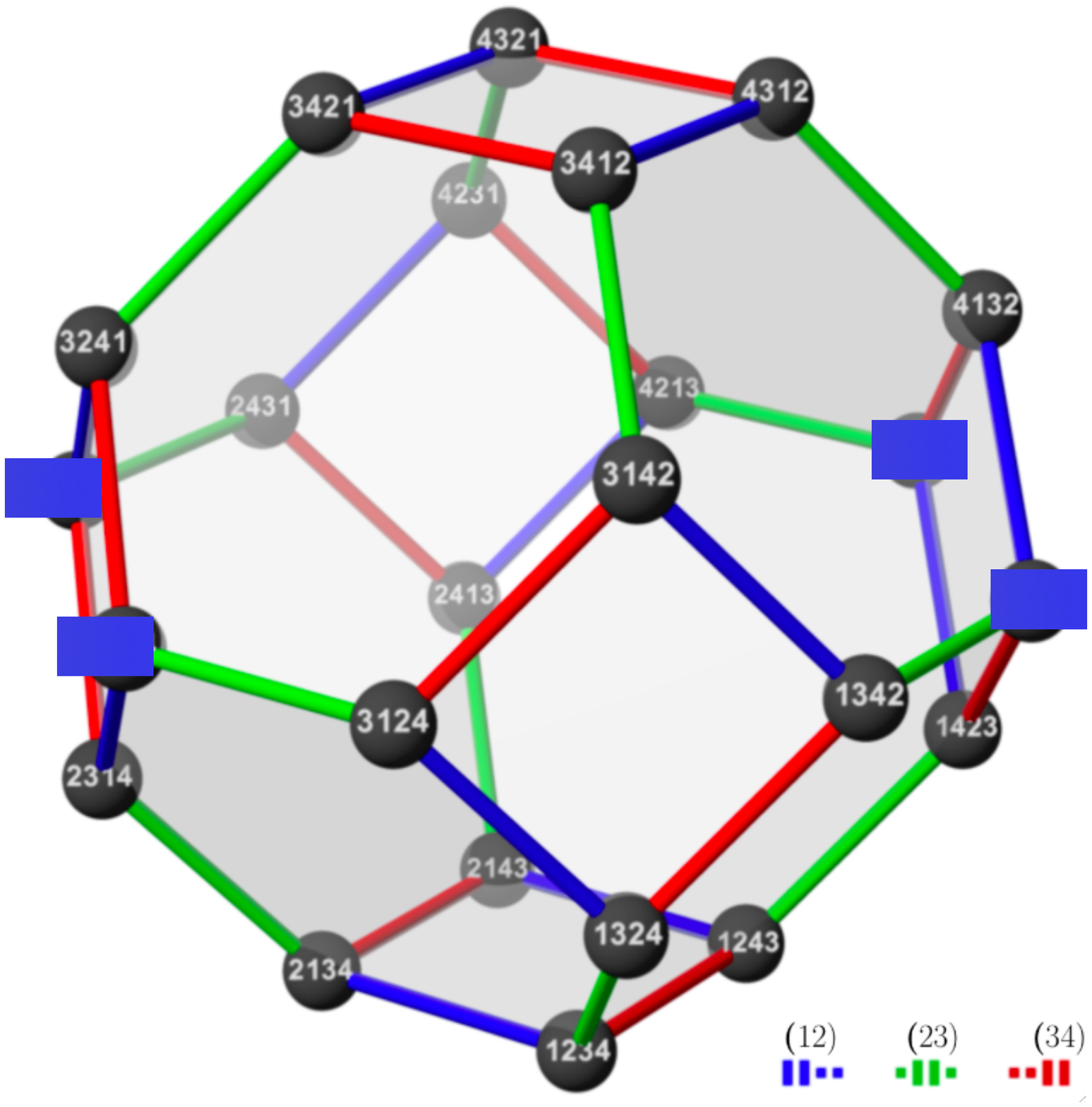}}
\put(-50,-104){{{\bf {Figure}} {\bf {12:}}
Permutahedron \& $ {\cal P}{}_{[4]}  $ Subset Elements }}
\end{picture}}
\nonumber
\label{fig:P-vm1}
$$
\vskip4.2in
The data contained in
${\bm {\cal A}}{}_{\ell x}[{\cal P}{}_{[a|4]},\, {\cal P}{}_{[b|4]} ] $ is presented below first in tabular 
\begin{equation}
\begin{tabular}{|c|ccccccccccc|} \hline
\diagbox{$\bm {{\cal P}{}_{[4]}}$}{$\bm {{\cal P}{}_{[4]}} $} &{~}& \,
$\langle 1432 \rangle$ & {~} & {~}$\langle 2341 \rangle$ & {~} & {~}$\langle 3214 \rangle$ &{~}& &$\langle 4123 \rangle$& & \\ \hline
$\langle 1432 \rangle$ &{~}& \,  0 & {~} & {~}6 & {~} & {~}4 &{~}& &2& & \\ \hline
$\langle 2341 \rangle$ &{~}& \,  6 & {~} & {~}0 & {~} & {~}2 &{~}& &4& & \\ \hline
$\langle 3214 \rangle$ &{~}& \,  4 & {~} & {~}2 & {~} & {~}0 &{~}& &6& & \\ \hline
$\langle 4123 \rangle$ &{~}& \,  2 & {~} & {~}4 & {~} & {~}6 &{~}& &0& & \\ \hline
\end{tabular}
\label{284}
\end{equation}  
\begin{center}
{{\bf Table 4:} $\{VM\}{}_1-\{VM\}{}_1 ~ {\rm {Two-Point ~  Correlator~ Values}}$}
 \end{center}
and in matrix form.
\be
{\bm {\cal A}}{}_{\ell x}[{\cal P}{}_{[a|4]},\, {\cal P}{}_{[b|4]} ] ~=~
\left[\begin{array}{cccc}
~0 & ~~6 &  ~~4  &  ~~2\\
~6 & ~~0 &  ~~2  &  ~~4\\
~4 & ~~2 &  ~~0  &  ~~6\\
~2 & ~~4 &  ~~6  &  ~~0
\end{array}\right]  {~~~~~~~~~~~}  {~~~~~~~~~~}
\label{eq:Mtrx4}
\ee

\newpage
\subsection{Intra-quartet  $\bm {{\cal P} {}_{[5]}} $ Correlators of the $\bm {{\mathbb{S}}{}_{4}}$ Permutahedron}
\label{sec:DeFNP5}

In this subsection, we illustrate the intra-quartet two-point correlators associated with $\bm {{\cal P} {}_{[5]}} $.  
The quartet members are indicated by brown quadrilaterals on some of the vertices of the permutahedron, the 
correlators are shown in tabular form, and finally this same data is presented as a matrix.

$$
\vCent
{\setlength{\unitlength}{1mm}
\begin{picture}(-20,0)
\put(-60,-120){\includegraphics[width=4.6in]{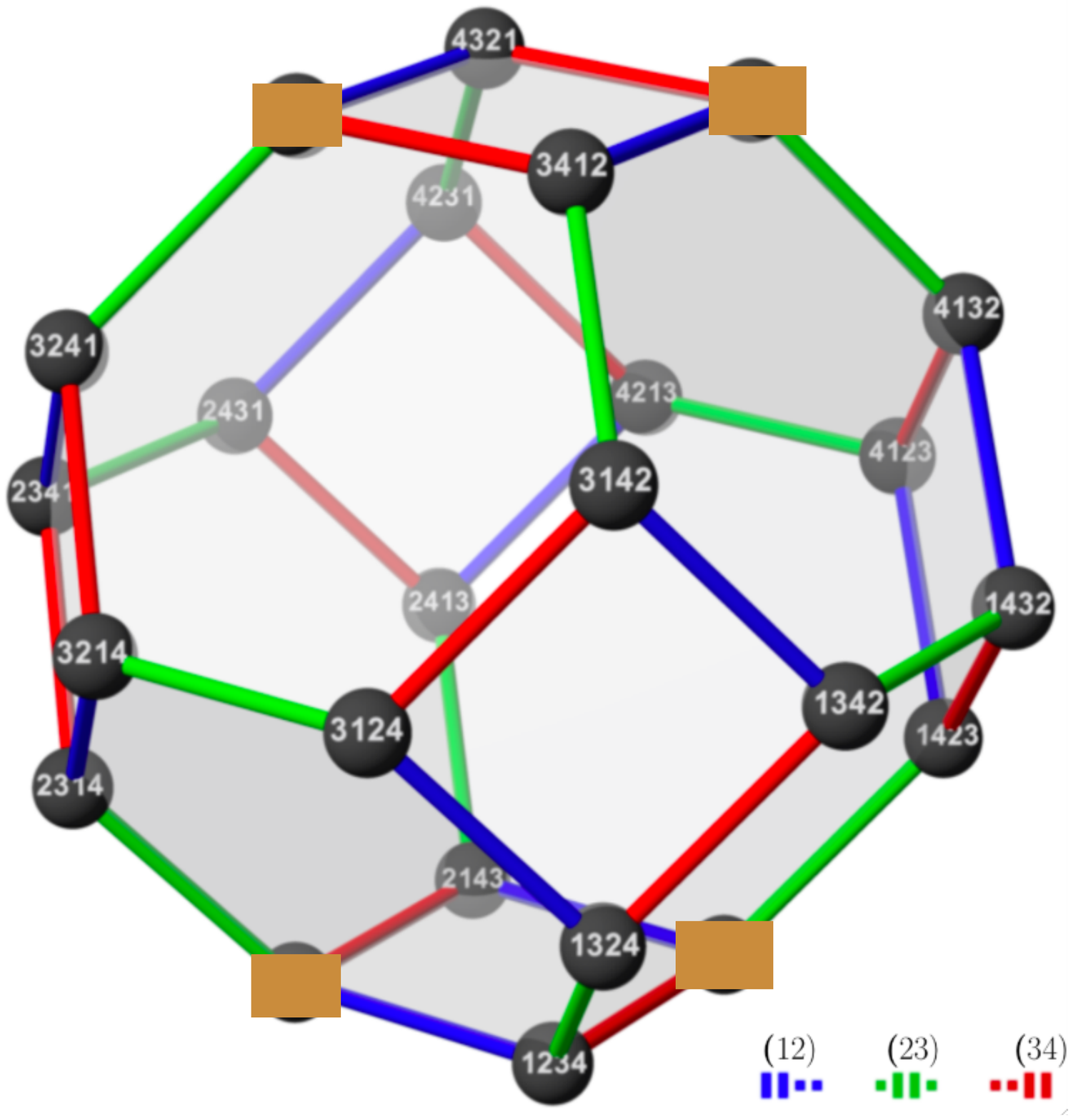}}
\put(-50,-104){{{\bf {Figure}} {\bf {13:}}
Permutahedron \& $ {\cal P}{}_{[5]}  $ Subset Elements }}
\end{picture}}
\nonumber
\label{fig:P-vm2}
$$
\vskip4.2in
The data contained in
${\bm {\cal A}}{}_{\ell x}[{\cal P}{}_{[a|5]},\, {\cal P}{}_{[b|5]} ] $ is presented below first in tabular 
\begin{equation}
\begin{tabular}{|c|ccccccccccc|} \hline
\diagbox{$\bm {{\cal P}{}_{[5]}}$}{$\bm {{\cal P}{}_{[5]}} $} &{~}& \,
$\langle 1243 \rangle$ & {~} & {~}$\langle 2134 \rangle$ & {~} & {~}$\langle 3421 \rangle$ &{~}& &$\langle 4312 \rangle$& & \\ \hline
$\langle 1243 \rangle$ &{~}& \,  0 & {~} & {~}2 & {~} & {~}6 &{~}& &4& & \\ \hline
$\langle 2134 \rangle$ &{~}& \,  2 & {~} & {~}0 & {~} & {~}4 &{~}& &6& & \\ \hline
$\langle 3421 \rangle$ &{~}& \,  6 & {~} & {~}4 & {~} & {~}0 &{~}& &2& & \\ \hline
$\langle 4312 \rangle$ &{~}& \,  4 & {~} & {~}6 & {~} & {~}2 &{~}& &0& & \\ \hline
\end{tabular}
\label{284}
\end{equation} 
\begin{center}
{{\bf Table 5:} $\{VM\}{}_2-\{VM\}{}_2 ~ {\rm {Two-Point ~  Correlator~ Values}}$}
\end{center}
and in matrix form.
\be
{\bm {\cal A}}{}_{\ell x}[{\cal P}{}_{[a|5]},\, {\cal P}{}_{[b|5]} ] ~=~
\left[\begin{array}{cccc}
~0 & ~~2 &  ~~6  &  ~~4\\
~2 & ~~0 &  ~~4  &  ~~6\\
~6 & ~~4 &  ~~0  &  ~~2\\
~4 & ~~6 &  ~~2  &  ~~0
\end{array}\right]  {~~~~~~~~~~~}  {~~~~~~~~~~}
\label{eq:Mtrx5}
\ee

\newpage
\subsection{Intra-quartet  $\bm {{\cal P} {}_{[6]}} $ Correlators of the $\bm {{\mathbb{S}}{}_{4}}$ Permutahedron}
\label{sec:DeFNP6}

In this subsection, we illustrate the intra-quartet two-point correlators associated with $\bm {{\cal P} {}_{[6]}} $.  
The quartet members are indicated by light blue quadrilaterals on some of the vertices of the permutahedron, the 
correlators are shown in tabular form, and finally this same data is presented as a matrix.

$$
\vCent
{\setlength{\unitlength}{1mm}
\begin{picture}(-20,0)
\put(-60,-120){\includegraphics[width=4.6in]{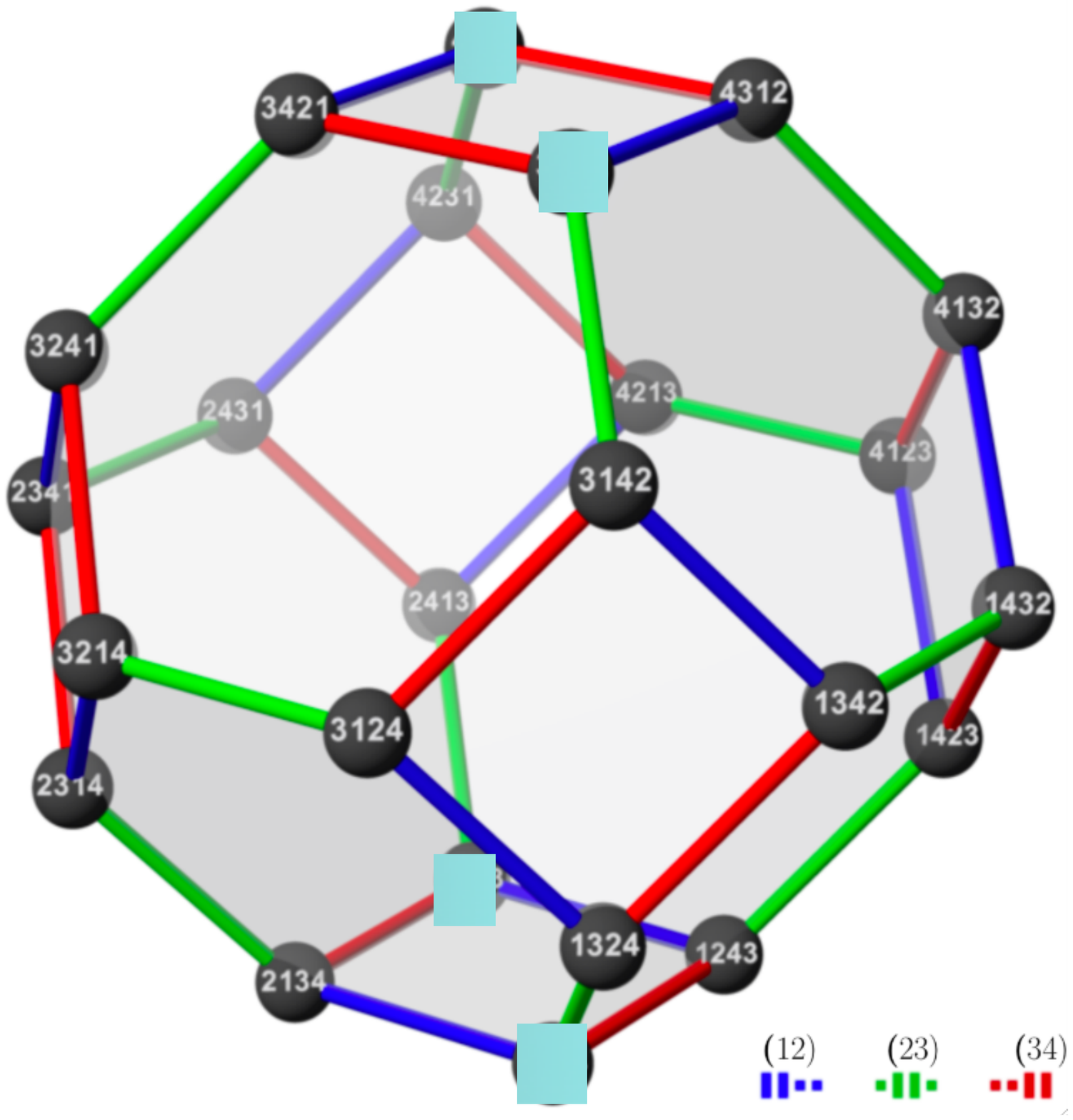}}
\put(-50,-104){{{\bf {Figure}} {\bf {14:}} Permutahedron \& $ {\cal P}{}_{[6]}  $ Subset Elements }}
\end{picture}}
\nonumber
\label{fig:P-vm3}
$$
\vskip4.2in
The data contained in
${\bm {\cal A}}{}_{\ell x}[{\cal P}{}_{[a|6]},\, {\cal P}{}_{[b|6]} ] $ is presented below first in tabular 
\begin{equation}
\begin{tabular}{|c|ccccccccccc|} \hline
\diagbox{$\bm {{\cal P}{}_{[6]}}$}{$\bm {{\cal P}{}_{[6]}} $} &{~}& \,
$\langle 1234 \rangle$ & {~} & {~}$\langle 2143 \rangle$ & {~} & {~}$\langle 3412 \rangle$ &{~}& &$\langle 4321 \rangle$& & \\ \hline
$\langle 1234 \rangle$ &{~}& \,  0 & {~} & {~}2 & {~} & {~}4 &{~}& &6& & \\ \hline
$\langle 2143 \rangle$ &{~}& \,  2 & {~} & {~}0 & {~} & {~}6 &{~}& &4& & \\ \hline
$\langle 3412 \rangle$ &{~}& \,  4 & {~} & {~}6 & {~} & {~}0 &{~}& &2& & \\ \hline
$\langle 4321 \rangle$ &{~}& \,  6 & {~} & {~}4 & {~} & {~}2 &{~}& &0& & \\ \hline
\end{tabular}
\label{284}
\end{equation}  
\begin{center}
{{\bf Table 6:} $\{VM\}{}_3-\{VM\}{}_3 ~ {\rm {Two-Point ~  Correlator~ Values}}$}
 \end{center}
and in matrix form.
\be
{\bm {\cal A}}{}_{\ell x}[{\cal P}{}_{[a|6]},\, {\cal P}{}_{[b|6]} ] ~=~
\left[\begin{array}{cccc}
~0 & ~~2 &  ~~4  &  ~~6\\
~2 & ~~0 &  ~~6  &  ~~4\\
~4 & ~~6 &  ~~0  &  ~~2\\
~6 & ~~4 &  ~~2  &  ~~0
\end{array}\right]  {~~~~~~~~~~~}  {~~~~~~~~~~}
\label{eq:Mtrx6}
\ee

\newpage
\section{ Inter-Quartet Results}
\label{sec:Calcs}

In this chapter we will give the explicit numerical result for the 
inter-quartet calculations without reference to any diagrams.

Given the six $\bm{\cal P}{}_{[A]}$ subsets, the minimal number of links it takes to travel from a given specified element $a$ in the set $\bm{\cal P}{}_{[A]}$ to a second specified element $b$ in the set $\bm{\cal P}{}_{[B]}$ where $A\ne B$ can form a table/matrix. From simply counting $6\times 5=30$, we can create in total 30 inter-quartet correlator matrices. 
Note that all these matrices satisfy the property that their transpose matrices have the same eigenvalues as themselves. Therefore we will only present half of them.


\be
\begin{tabular}{|c|ccccccccccc|} \hline
\diagbox{$\bm {{\cal P}{}_{[6]}}$}{$\bm {{\cal P}{}_{[1]}} $} &{~}& \,  
$\langle 1423 \rangle$ & {~} & {~}$\langle 2314 \rangle$ & {~} & {~}$\langle 3241 \rangle$ &{~}& &$\langle 4132 \rangle$& & \\ \hline
$\langle 1234 \rangle$ &{~}& \,  2 & {~} & {~}2 & {~} & {~}4 &{~}& &4& & \\ \hline
$\langle 2143 \rangle$ &{~}& \,  2 & {~} & {~}2 & {~} & {~}4 &{~}& &4& & \\ \hline
$\langle 3412 \rangle$ &{~}& \,  4 & {~} & {~}4 & {~} & {~}2 &{~}& &2& & \\ \hline
$\langle 4321 \rangle$ &{~}& \,  4 & {~} & {~}4 & {~} & {~}2 &{~}& &2& & \\ \hline
\end{tabular}
\label{eq:T1}
\ee 
\begin{center}
{{\bf Table 7:} $\{{\cal P}\}{}_6-\{ {\cal P}\}{}_1 ~ {\rm {Two-Point ~  Correlator~ Values}}$}
 \end{center}
Matrix Eigenvalues: $\{12,\, -4,\, 0,\, 0\}$  $ ~~~~$ Matrix Trace = 8


\begin{equation}
\begin{tabular}{|c|ccccccccccc|} \hline
\diagbox{$\bm {{\cal P}{}_{[6]}}$}{$\bm {{\cal P}{}_{[2]}} $} &{~}& \,  
$\langle 1342 \rangle$ & {~} & {~}$\langle 2431 \rangle$ & {~} & {~}$\langle 3124 \rangle$ &{~}& &$\langle 4213 \rangle$& & \\ \hline
$\langle 1234 \rangle$ &{~}& \,  2 & {~} & {~}4 & {~} & {~}2 &{~}& &4& & \\ \hline
$\langle 2143 \rangle$ &{~}& \,  4 & {~} & {~}2 & {~} & {~}4 &{~}& &2& & \\ \hline
$\langle 3412 \rangle$ &{~}& \,  2 & {~} & {~}4 & {~} & {~}2 &{~}& &4& & \\ \hline
$\langle 4321 \rangle$ &{~}& \,  4 & {~} & {~}2 & {~} & {~}4 &{~}& &2& & \\ \hline
\end{tabular}
\label{eq:T2}
\end{equation} 
\begin{center}
{{\bf Table 8:} $\{{\cal P}\}{}_6-\{ {\cal P}\}{}_2 ~ {\rm {Two-Point ~  Correlator~ Values}}$}
 \end{center}
Matrix Eigenvalues: $\{12,\, -4,\, 0,\, 0\}$  $ ~~~~$ Matrix Trace = 8


\begin{equation}
\begin{tabular}{|c|ccccccccccc|} \hline
\diagbox{$\bm {{\cal P}{}_{[6]}}$}{$\bm {{\cal P}{}_{[3]}} $} &{~}& \,  
$\langle 1324 \rangle$ & {~} & {~}$\langle 2413 \rangle$ & {~} & {~}$\langle 3142 \rangle$ &{~}& &$\langle 4231 \rangle$& & \\ \hline
$\langle 1234 \rangle$ &{~}& \,  1 & {~} & {~}3 & {~} & {~}3 &{~}& &5& & \\ \hline
$\langle 2143 \rangle$ &{~}& \,  3 & {~} & {~}1 & {~} & {~}5 &{~}& &3& & \\ \hline
$\langle 3412 \rangle$ &{~}& \,  3 & {~} & {~}5 & {~} & {~}1 &{~}& &3& & \\ \hline
$\langle 4321 \rangle$ &{~}& \,  5 & {~} & {~}3 & {~} & {~}3 &{~}& &1& & \\ \hline
\end{tabular}
\label{eq:T3}
\end{equation} 
\begin{center}
{{\bf Table 9:} $\{{\cal P}\}{}_6-\{ {\cal P}\}{}_3 ~ {\rm {Two-Point ~  Correlator~ Values}}$}
 \end{center}
Matrix Eigenvalues:  $\{12,\, -4,\, -4,\, 0\}$
 $ ~~~~$ Matrix Trace = 4
 \newpage


\begin{equation}
\begin{tabular}{|c|ccccccccccc|} \hline
\diagbox{$\bm {{\cal P}{}_{[6]}}$}{$\bm {{\cal P}_{[4]}} $} &{~}& \,  
$\langle 1432 \rangle$ & {~} & {~}$\langle 2341 \rangle$ & {~} & {~}$\langle 3214 \rangle$ &{~}& &$\langle 4123 \rangle$& & \\ \hline
$\langle 1234 \rangle$ &{~}& \,  3 & {~} & {~}3 & {~} & {~}3 &{~}& &3& & \\ \hline
$\langle 2143 \rangle$ &{~}& \,  3 & {~} & {~}3 & {~} & {~}3 &{~}& &3& & \\ \hline
$\langle 3412 \rangle$ &{~}& \,  3 & {~} & {~}3 & {~} & {~}3 &{~}& &3& & \\ \hline
$\langle 4321 \rangle$ &{~}& \,  3 & {~} & {~}3 & {~} & {~}3 &{~}& &3& & \\ \hline
\end{tabular}
\label{eq:T4}
\end{equation} 
\begin{center}
{{\bf Table 10:} $\{{\cal P}\}{}_6-\{ {\cal P}\}{}_4 ~ {\rm {Two-Point ~  Correlator~ Values}}$}
 \end{center}
Matrix Eigenvalues: $\{12,\, 0,\, 0,\, 0\}$
 $ ~~~~$ Matrix Trace = 12

\begin{equation}
\begin{tabular}{|c|ccccccccccc|} \hline
\diagbox{$\bm {{\cal P}{}_{[6]}}$}{$\bm {{\cal P}{}_{[5]}} $} &{~}& \,  
$\langle 1243 \rangle$ & {~} & {~}$\langle 2134 \rangle$ & {~} & {~}$\langle 3421 \rangle$ &{~}& &$\langle 4312 \rangle$& & \\ \hline
$\langle 1234 \rangle$ &{~}& \,  1 & {~} & {~}1 & {~} & {~}5 &{~}& &5& & \\ \hline
$\langle 2143 \rangle$ &{~}& \,  1 & {~} & {~}1 & {~} & {~}5 &{~}& &5& & \\ \hline
$\langle 3412 \rangle$ &{~}& \,  5 & {~} & {~}5 & {~} & {~}1 &{~}& &1& & \\ \hline
$\langle 4321 \rangle$ &{~}& \,  5 & {~} & {~}5 & {~} & {~}1 &{~}& &1& & \\ \hline
\end{tabular}
\label{eq:T5}
\end{equation} 
\begin{center}
{{\bf Table 11:} $\{{\cal P}\}{}_6-\{ {\cal P}\}{}_5 ~ {\rm {Two-Point ~  Correlator~ Values}}$}
 \end{center}
Matrix Eigenvalues:  $\{12,\, -8,\, 0,\, 0\}$    $ ~~~~$ Matrix Trace = 4


\begin{equation}
\begin{tabular}{|c|ccccccccccc|} \hline
\diagbox{$\bm {{\cal P}{}_{[5]}}$}{$\bm {{\cal P}{}_{[1]}} $} &{~}& \,  
$\langle 1423 \rangle$ & {~} & {~}$\langle 2314 \rangle$ & {~} & {~}$\langle 3241 \rangle$ &{~}& &$\langle 4132 \rangle$& & \\ \hline
$\langle 1243 \rangle$ &{~}& \,  1 & {~} & {~}3 & {~} & {~}5 &{~}& &3& & \\ \hline
$\langle 2134 \rangle$ &{~}& \,  3 & {~} & {~}1 & {~} & {~}3 &{~}& &5& & \\ \hline
$\langle 3421 \rangle$ &{~}& \,  5 & {~} & {~}3 & {~} & {~}1 &{~}& &3& & \\ \hline
$\langle 4312 \rangle$ &{~}& \,  3 & {~} & {~}5 & {~} & {~}3 &{~}& &1& & \\ \hline
\end{tabular}
\label{eq:T6}
\end{equation} 
\begin{center}
{{\bf Table 12:} $\{{\cal P}\}{}_5-\{ {\cal P}\}{}_1 ~ {\rm {Two-Point ~  Correlator~ Values}}$}
 \end{center}
Matrix Eigenvalues: $\{12,\, -4,\, -4,\, 0\}$  $ ~~~~$ Matrix Trace = 4


\begin{equation}
\begin{tabular}{|c|ccccccccccc|} \hline
\diagbox{$\bm {{\cal P}{}_{[5]}}$}{$\bm {{\cal P}{}_{[2]}} $} &{~}& \,  
$\langle 1342 \rangle$ & {~} & {~}$\langle 2431 \rangle$ & {~} & {~}$\langle 3124 \rangle$ &{~}& &$\langle 4213 \rangle$& & \\ \hline
$\langle 1243 \rangle$ &{~}& \,  3 & {~} & {~}3 & {~} & {~}3 &{~}& &3& & \\ \hline
$\langle 2134 \rangle$ &{~}& \,  3 & {~} & {~}3 & {~} & {~}3 &{~}& &3& & \\ \hline
$\langle 3421 \rangle$ &{~}& \,  3 & {~} & {~}3 & {~} & {~}3 &{~}& &3& & \\ \hline
$\langle 4312 \rangle$ &{~}& \,  3 & {~} & {~}3 & {~} & {~}3 &{~}& &3& & \\ \hline
\end{tabular}
\label{eq:T7}
\end{equation} 
\begin{center}
{{\bf Table 13:} $\{{\cal P}\}{}_5-\{ {\cal P}\}{}_2 ~ {\rm {Two-Point ~  Correlator~ Values}}$}
 \end{center}
Matrix Eigenvalues: $\{12,\, 0,\, 0,\, 0\}$
 $ ~~~~$ Matrix Trace = 12


\begin{equation}
\begin{tabular}{|c|ccccccccccc|} \hline
\diagbox{$\bm {{\cal P}{}_{[5]}}$}{$\bm {{\cal P}{}_{[3]}} $} &{~}& \,  
$\langle 1324 \rangle$ & {~} & {~}$\langle 2413 \rangle$ & {~} & {~}$\langle 3142 \rangle$ &{~}& &$\langle 4231 \rangle$& & \\ \hline
$\langle 1243 \rangle$ &{~}& \,  2 & {~} & {~}2 & {~} & {~}4 &{~}& &4& & \\ \hline
$\langle 2134 \rangle$ &{~}& \,  2 & {~} & {~}2 & {~} & {~}4 &{~}& &4& & \\ \hline
$\langle 3421 \rangle$ &{~}& \,  4 & {~} & {~}4 & {~} & {~}2 &{~}& &2& & \\ \hline
$\langle 4312 \rangle$ &{~}& \,  4 & {~} & {~}4 & {~} & {~}2 &{~}& &2& & \\ \hline
\end{tabular}
\label{eq:T8}
\end{equation} 
\begin{center}
{{\bf Table 14:} $\{{\cal P}\}{}_5-\{ {\cal P}\}{}_3 ~ {\rm {Two-Point ~  Correlator~ Values}}$}
 \end{center}
Matrix Eigenvalues: $\{12,\, -4,\, 0,\, 0\}$  $ ~~~~$ Matrix Trace = 8


\begin{equation}
\begin{tabular}{|c|ccccccccccc|} \hline
\diagbox{$\bm {{\cal P}{}_{[5]}}$}{$\bm {{\cal P}{}_{[4]}} $} &{~}& \,  
$\langle 1432 \rangle$ & {~} & {~}$\langle 2341 \rangle$ & {~} & {~}$\langle 3214 \rangle$ &{~}& &$\langle 4123 \rangle$& & \\ \hline
$\langle 1243 \rangle$ &{~}& \,  2 & {~} & {~}4 & {~} & {~}4 &{~}& &2& & \\ \hline
$\langle 2134 \rangle$ &{~}& \,  4 & {~} & {~}2 & {~} & {~}2 &{~}& &4& & \\ \hline
$\langle 3421 \rangle$ &{~}& \,  4 & {~} & {~}2 & {~} & {~}2 &{~}& &4& & \\ \hline
$\langle 4312 \rangle$ &{~}& \,  2 & {~} & {~}4 & {~} & {~}4 &{~}& &2& & \\ \hline
\end{tabular}
\label{eq:T9}
\end{equation} 
\begin{center}
{{\bf Table 15:} $\{{\cal P}\}{}_5-\{ {\cal P}\}{}_4 ~ {\rm {Two-Point ~  Correlator~ Values}}$}
 \end{center}
Matrix Eigenvalues: $\{12,\, -4,\, 0,\, 0\}$  $ ~~~~$ Matrix Trace = 8


\begin{equation}
\begin{tabular}{|c|ccccccccccc|} \hline
\diagbox{$\bm {{\cal P}{}_{[4]}}$}{$\bm {{\cal P}{}_{[1]}} $} &{~}& \,  
$\langle 1423 \rangle$ & {~} & {~}$\langle 2314 \rangle$ & {~} & {~}$\langle 3241 \rangle$ &{~}& &$\langle 4132 \rangle$& & \\ \hline
$\langle 1432 \rangle$ &{~}& \,  1 & {~} & {~}5 & {~} & {~}5 &{~}& &1& & \\ \hline
$\langle 2341 \rangle$ &{~}& \,  5 & {~} & {~}1 & {~} & {~}1 &{~}& &5& & \\ \hline
$\langle 3214 \rangle$ &{~}& \,  5 & {~} & {~}1 & {~} & {~}1 &{~}& &5& & \\ \hline
$\langle 4123 \rangle$ &{~}& \,  1 & {~} & {~}5 & {~} & {~}5 &{~}& &1& & \\ \hline
\end{tabular}
\label{eq:T10}
\end{equation} 
\begin{center}
{{\bf Table 16:} $\{{\cal P}\}{}_4-\{ {\cal P}\}{}_1 ~ {\rm {Two-Point ~  Correlator~ Values}}$}
 \end{center}
Matrix Eigenvalues: $\{12,\, -8,\, 0,\, 0\}$  $ ~~~~$ Matrix Trace = 4


\begin{equation}
\begin{tabular}{|c|ccccccccccc|} \hline
\diagbox{$\bm {{\cal P}{}_{[4]}}$}{$\bm {{\cal P}{}_{[2]}} $} &{~}& \,  
$\langle 1342 \rangle$ & {~} & {~}$\langle 2431 \rangle$ & {~} & {~}$\langle 3124 \rangle$ &{~}& &$\langle 4213 \rangle$& & \\ \hline
$\langle 1432 \rangle$ &{~}& \,  1 & {~} & {~}5 & {~} & {~}3 &{~}& &3& & \\ \hline
$\langle 2341 \rangle$ &{~}& \,  5 & {~} & {~}1 & {~} & {~}3 &{~}& &3& & \\ \hline
$\langle 3214 \rangle$ &{~}& \,  3 & {~} & {~}3 & {~} & {~}1 &{~}& &5& & \\ \hline
$\langle 4123 \rangle$ &{~}& \,  3 & {~} & {~}3 & {~} & {~}5 &{~}& &1& & \\ \hline
\end{tabular}
\label{eq:T11}
\end{equation} 
\begin{center}
{{\bf Table 17:} $\{{\cal P}\}{}_4-\{ {\cal P}\}{}_2 ~ {\rm {Two-Point ~  Correlator~ Values}}$}
 \end{center}
Matrix Eigenvalues: $\{12,\, -4,\, -4,\, 0\}$  $ ~~~~$ Matrix Trace = 4


\begin{equation}
\begin{tabular}{|c|ccccccccccc|} \hline
\diagbox{$\bm {{\cal P}{}_{[4]}}$}{$\bm {{\cal P}{}_{[3]}} $} &{~}& \,  
$\langle 1324 \rangle$ & {~} & {~}$\langle 2413 \rangle$ & {~} & {~}$\langle 3142 \rangle$ &{~}& &$\langle 4231 \rangle$& & \\ \hline
$\langle 1432 \rangle$ &{~}& \,  2 & {~} & {~}4 & {~} & {~}2 &{~}& &4& & \\ \hline
$\langle 2341 \rangle$ &{~}& \,  4 & {~} & {~}2 & {~} & {~}4 &{~}& &2& & \\ \hline
$\langle 3214 \rangle$ &{~}& \,  2 & {~} & {~}4 & {~} & {~}2 &{~}& &4& & \\ \hline
$\langle 4123 \rangle$ &{~}& \,  4 & {~} & {~}2 & {~} & {~}4 &{~}& &2& & \\ \hline
\end{tabular}
\label{eq:T12}
\end{equation} 
\begin{center}
{{\bf Table 18:} $\{{\cal P}\}{}_4-\{ {\cal P}\}{}_3 ~ {\rm {Two-Point ~  Correlator~ Values}}$}
 \end{center}
Matrix Eigenvalues: $\{12,\, -4,\, 0,\, 0\}$  $ ~~~~$ Matrix Trace = 8


\begin{equation}
\begin{tabular}{|c|ccccccccccc|} \hline
\diagbox{$\bm {{\cal P}{}_{[3]}}$}{$\bm {{\cal P}{}_{[1]}} $} &{~}& \,  
$\langle 1423 \rangle$ & {~} & {~}$\langle 2314 \rangle$ & {~} & {~}$\langle 3241 \rangle$ &{~}& &$\langle 4132 \rangle$& & \\ \hline
$\langle 1324 \rangle$ &{~}& \,  3 & {~} & {~}3 & {~} & {~}3 &{~}& &3& & \\ \hline
$\langle 2413 \rangle$ &{~}& \,  3 & {~} & {~}3 & {~} & {~}3 &{~}& &3& & \\ \hline
$\langle 3142 \rangle$ &{~}& \,  3 & {~} & {~}3 & {~} & {~}3 &{~}& &3& & \\ \hline
$\langle 4231 \rangle$ &{~}& \,  3 & {~} & {~}3 & {~} & {~}3 &{~}& &3& & \\ \hline
\end{tabular}
\label{eq:T13}
\end{equation} 
\begin{center}
{{\bf Table 19:} $\{{\cal P}\}{}_3-\{ {\cal P}\}{}_1 ~ {\rm {Two-Point ~  Correlator~ Values}}$}
 \end{center}
Matrix Eigenvalues:  $\{12,\, 0,\, 0,\, 0\}$
 $ ~~~~$ Matrix Trace = 12 


\begin{equation}
\begin{tabular}{|c|ccccccccccc|} \hline
\diagbox{$\bm {{\cal P}{}_{[3]}}$}{$\bm {{\cal P}{}_{[2]}} $} &{~}& \,  
$\langle 1342 \rangle$ & {~} & {~}$\langle 2431 \rangle$ & {~} & {~}$\langle 3124 \rangle$ &{~}& &$\langle 4213 \rangle$& & \\ \hline
$\langle 1324 \rangle$ &{~}& \,  1 & {~} & {~}5 & {~} & {~}1 &{~}& &5& & \\ \hline
$\langle 2413 \rangle$ &{~}& \,  5 & {~} & {~}1 & {~} & {~}5 &{~}& &1& & \\ \hline
$\langle 3142 \rangle$ &{~}& \,  1 & {~} & {~}5 & {~} & {~}1 &{~}& &5& & \\ \hline
$\langle 4231 \rangle$ &{~}& \,  5 & {~} & {~}1 & {~} & {~}5 &{~}& &1& & \\ \hline
\end{tabular}
\label{eq:T14}
\end{equation} 
\begin{center}
{{\bf Table 20:} $\{{\cal P}\}{}_3-\{ {\cal P}\}{}_2 ~ {\rm {Two-Point ~  Correlator~ Values}}$}
 \end{center}
Matrix Eigenvalues: $\{12,\, -4,\, 0,\, 0\}$  $ ~~~~$ Matrix Trace = 4


\begin{equation}
\begin{tabular}{|c|ccccccccccc|} \hline
\diagbox{$\bm {{\cal P}{}_{[2]}}$}{$\bm {{\cal P}{}_{[1]}} $} &{~}& \,  
$\langle 1423 \rangle$ & {~} & {~}$\langle 2314 \rangle$ & {~} & {~}$\langle 3241 \rangle$ &{~}& &$\langle 4132 \rangle$& & \\ \hline
$\langle 1342 \rangle$ &{~}& \,  2 & {~} & {~}4 & {~} & {~}4 &{~}& &2& & \\ \hline
$\langle 2431 \rangle$ &{~}& \,  4 & {~} & {~}2 & {~} & {~}2 &{~}& &4& & \\ \hline
$\langle 3124 \rangle$ &{~}& \,  4 & {~} & {~}2 & {~} & {~}2 &{~}& &4& & \\ \hline
$\langle 4213 \rangle$ &{~}& \,  2 & {~} & {~}4 & {~} & {~}4 &{~}& &2& & \\ \hline
\end{tabular}
\label{eq:T15}
\end{equation} 
\begin{center}
{{\bf Table 21:} $\{{\cal P}\}{}_2-\{ {\cal P}\}{}_1 ~ {\rm {Two-Point ~  Correlator~ Values}}$}
 \end{center}
Matrix Eigenvalues: $\{12,\, -4,\, 0,\, 0\}$  $ ~~~~$ Matrix Trace = 8
$$~~$$

\newpage
\section{20,736 Permutahedronic ``Correlators''
Via Eigenvalue Equivalence Classes}
\label{sec:DeFNs}

For an adinkra, if we remove all dashing considerations, the remaining graph is called the chromotopology of the adinkra. In the context of chromotopology, there's only one ``seed'' adinkra required to generate all four-color minimal adinkras: the Vierergruppe\footnote{The Vierergruppe or Klein 4-group corresponds $\bm {{\cal P}{}_{[6]}}. $ } \cite{Gates:2017eui}. 
The total number of four-color minimal adinkras carrying different chromotopology is ${\rm ord}(\mathbb{S}_3)\times {\rm ord}(\mathbb{S}_4)=144$. More generally, as shown in \cite{Gates:2017eui}, left cosets of the Vierergruppe via $S_3$ generate all elements of $S_4$ (which are our $\textbf{L}_{I}$), and the Coxeter group $B C_{4}^{\pm a \mu A}=\pm H^{a} S_{3}^{\mu} \mathcal{V}^{A}$, where  $H^{a}$ just corresponds to a sign flip and $\mathcal{V}^{A}$ is the Vierergruppe.

In order to generate these 144 adinkras, we can start from the six $\bm{\cal P}_{[i]}$ sets we presented before. Each of them belongs to a family of adinkras that are generated by performing color permutations. 
For example, look at ${\cal P}_{[1]} = \{\langle1423\rangle, \langle2314\rangle,\langle3241\rangle,\langle4132\rangle\}$. Although it looks like a set, the order of matrices does matter! We can assign ${\bm {\rL}}$-matrices for this adinkra as
\begin{equation}
    {\bm {\rL}}_1~=~\langle1423\rangle~,~ {\bm {\rL}}_2~=~\langle2314\rangle~,~{\bm {\rL}}_3~=~\langle3241\rangle~,~{\bm {\rL}}_4~=~\langle4132\rangle
\end{equation}
Note that the index $\rI$ attached on L-matrices ${\bm {\rL}}{}_{{}_{\rI}}$ labels the color. By doing color permutations, there are in total $4!=24$ descendant adinkras. One example is to switch color 1 and color 2, i.e.
\begin{equation}
    {\bm {\rL}}_1~=~\langle2314\rangle~,~ {\bm {\rL}}_2~=~\langle1423\rangle~,~{\bm {\rL}}_3~=~\langle3241\rangle~,~{\bm {\rL}}_4~=~\langle4132\rangle
\end{equation}

We can label these six families as $\{\bm{\cal P}_{[i]}\}$, where each family has $4!$ adinkras. Clearly there's no intersection between any two families. 

Then, we can study all $144\times 144=20,736$ distance matrices and classify them by eigenvalues. In order to give a clear presentation as well as maximize computational efficiency, these 20,736 matrices can be classified based on their origins as below. 

(1.) $3,456=4!\times4!\times6$ matrices generated by two adinkras from the same family $\{\bm{\cal P}_{[i]}\}$. 

(2.) $17280=4!\times4!\times2\times15$ matrices generated by two adinkras from two different families $\{\bm{\cal P}_{[i]}\}$ and $\{\bm{\cal P}_{[j]}\}$. 
Note that results in Chapter \ref{sec:Calcs} implies that this class can be further divided in four subclasses: 

(2.1) $(i,j) = (1,2), (1,6), (2,6), (3,4), (3,5), (4,5)$

(2.2) $(i,j) = (2,3), (5,6), (1,4)$

(2.3) $(i,j) = (1,3), (2,5), (4,6)$

(2.4) $(i,j) = (1,5), (2,4), (3,6)$

It turns out that many of these 20,736 distance matrices have the same eigenvalues. Finally, there are only 15 unique sets of eigenvalues. Table 22 shows the summary of eigenvalue equivalent classes of 20,736 distance matrices. You can find intermediate tables obtained from class (1) to class (2.4) in Appendix \ref{sec:classes}.

\newpage
\begin{table}[htp!]
    \centering
    \begin{tabular}{|c|c|c|}
    \hline
       Eigenvalues  & \# of Matrices & Trace \\\hline\hline
        ${12, 8, 4, 0}$ & 144 & 24 \\\hline
        ${12, 8, 0, 0}$ & 864 & 20 \\\hline
        ${12, 4, 4,0}$ & 144 & 20 \\\hline
        ${12, 4, 0, 0}$ & 2,016 & 16 \\\hline
        ${12, 8, -4, 0}$ & 144 & 16 \\\hline
         ${12, 0, 0, 0}$ & 12,672  & 12 \\\hline
         ${12, -4, 4, 0}$ & 576 & 12 \\\hline
          ${12, 4i, -4i, 0}$ & 288 & 12 \\\hline
         ${12, -4\sqrt{2}, 4\sqrt{2}, 0}$  & 288 & 12 \\\hline
        ${12, -4i\sqrt{2}, 4i\sqrt{2}, 0}$  & 288 & 12 \\\hline
        ${12, -4, 0, 0}$ & 2,016 & 8 \\\hline
         ${12, -8, 4, 0}$ & 144 &  8\\\hline
         $ {12, -8, 0, 0}$ & 864 & 4 \\\hline
          ${12, -4, -4, 0}$ & 144 & 4 \\\hline
           ${12, -8, -4, 0}$ & 144 & 0 \\\hline      
    \end{tabular}
    \label{tab:sum_class}
\end{table}
\begin{center}
   {{\bf Table 22:} Eigenvalue Equivalent Classes of 20,736 Distance Matrices}
 \end{center}
 
In Chapter \ref{sec:DeFNP1} and \ref{sec:Calcs}, we discussed a small subset of this analysis. Let's discuss one more explicit example: consider an adinkra A belong in family $\{\bm{\cal P}_{[1]}\}$ with ${\bm {\rL}}$-matrices
\begin{equation}
    {\bm {\rL}}_{1,A}~=~\langle2314\rangle~,~ {\bm {\rL}}_{2,A}~=~\langle3241\rangle~,~{\bm {\rL}}_{3,A}~=~\langle1423\rangle~,~{\bm {\rL}}_{4,A}~=~\langle4132\rangle
\end{equation}
and adinkra B belong in family $\{\bm{\cal P}_{[1]}\}$ with ${\bm {\rL}}$-matrices
\begin{equation}
    {\bm {\rL}}_{1,B}~=~\langle2314\rangle~,~ {\bm {\rL}}_{2,B}~=~\langle1423\rangle~,~{\bm {\rL}}_{3,B}~=~\langle3241\rangle~,~{\bm {\rL}}_{4,B}~=~\langle4132\rangle
\end{equation}
Then the distance matrix corresponding to these two adinkras is 
\begin{equation}
\begin{tabular}{|c|ccccccccccc|} \hline
\diagbox{$A$}{$B$} &{~}& \,  
$\langle 2314 \rangle$ & {~} & {~}$\langle 1423 \rangle$ & {~} & {~}$\langle 3241 \rangle$ &{~}& &$\langle 4132 \rangle$& & \\ \hline
$\langle 2314 \rangle$ &{~}& \,  0 & {~} & {~}4 & {~} & {~}2 &{~}& &6& & \\ \hline
$\langle 3241 \rangle$ &{~}& \,  2 & {~} & {~}6 & {~} & {~}0 &{~}& &4& & \\ \hline
$\langle 1423 \rangle$ &{~}& \,  4 & {~} & {~}0 & {~} & {~}6 &{~}& &2& & \\ \hline
$\langle 4132 \rangle$ &{~}& \,  6 & {~} & {~}2 & {~} & {~}4 &{~}& &0& & \\ \hline
\end{tabular}
\label{eq:example}
\end{equation}
whose eigenvalues are $\{12,-4\sqrt{2},4\sqrt{2},0\}$.

\newpage 
\section{The View Beyond 4D, \texorpdfstring{$\cal N$}{N} = 1 SUSY}
\label{sec:BeYnd}

In this section we wish to connect the current presentation relating the
geometry of the orbits of elements described by the permutahedron to previous
discussions \cite{GRana1,GRana2,GHIM} which go beyond the exemplary
case of ${\mathbb{BC}}{}_{4}$ used in this work.

The faces that minimally connect all the elements contained in any single quartet are 
squares. Among the subsets $\bm {{\cal P}{}_{[1]}}$, $\cdots$, $\bm {{\cal P}{}_{[6]}}$,
there is an element that has the minimum values with respect to lexicographical ordering.  
This is the identity element $()$ and we note $()$ $\in$ $\bm {{\cal P}{}_{[6]}}$. In
the following, we will go beyond the case related to 4D, $\cal N $ = 1 SUSY to
discuss similar structures that have been found for arbitrary values of $\cal N$.
Let us recall that $\bm {{\cal P}{}_{[6]}}$ is ``Klein’s Vierergruppe" and we will
concentrate on this generalization for arbitrary $\cal N$.

Already in the work of \cite{GRana1,GRana2} the existence of a recursive procedure
was demonstrated for how to construct the analog of the Vierergruppe that is needed
for all values of $\cal N$.  We will present this recursion in an appendix. In
the work of \cite{GHIM} the case for $\cal N$ = 8, the analogue of the elements of
the Vierergruppe were presented as $\bm {{\cal P}_{\Hat I}} $ where explicitly we see
\be{
\begin{aligned}
{\cal P}_1&=( )~~~,~~~
{\cal P}_2=(1 2)(3 4)(5 6)(7 8)~~~,~~~
{\cal P}_3=(1 3)(2 4)(5 7)(6 8)~~~,~~~
{\cal P}_4=(1 4)(2 3)(5 8)(6 7)~~~, \\
{\cal P}_5&=(1 5)(2 6)(3 7)(4 8)~~~,~~~
{\cal P}_6=(1 6)(2 5)(3 8)(4 7)~~~,~~~
{\cal P}_7=(1 7)(2 8)(3 5)(4 6)~~~,~~\\
{\cal P}_8&=(1 8)(2 7)(3 6)(4 5)~~~. 
\end{aligned}
}
\label{eq:N8ps}
\ee
as the index $\bm {{\Hat I}}$ ranges from 1, $\dots$, 8.

Looking back at Fig.\ 9 the Vierergruppe consists of the permutation elements $\langle 1234 \rangle$, 
$\langle 2134 \rangle$, $\langle 1243 \rangle$, and $\langle 2143 \rangle$, all of which occur at the 
`base'' of the permutahedron related to 4D, $\cal N$ = 1 minimal supersymmetry representations.  By 
comparison, the permutation elements of ${\mathbb S}{}_8 $ that appear in Eq.\ (\ref{eq:N8ps}) must 
occur at the ``base'' of the permutahedron related to 4D, $\cal N$ = 2 minimal supersymmetry 
representations.  Application of the pair-wise adjacent permutations $(23)$, $(34)$, $(45)$,
$(56)$, $(67)$, and $(78)$ to all the permutation elements that appear in Eq.\ (\ref{eq:N8ps})
will lead to the complete permutahedron associated with 4D, $\cal N$ = 2 minimal supersymmetry 
representations.

The size of the permutation matrices ``jumps" by a factor of the $\cal N$ values belong to sequence
of numbers given by
\be
    1 , ~ 2, ~ 4, ~ 8, ~ 9, ~ 10, ~12, ~ 16, ~17, ~18, ~20, ~25, \dots
\ee
We borrow language from nuclear physics
\footnote{See the webpage at https://en.wikipedia.org/wiki/Magic$_-$number$_-$(physics) on-line.} 
and call these ``magical'' values of $\cal N$. Once the matrix representation of the permutations is known, it is a straightforward by tedious matter to convert the results into either $\langle  \rangle$-notation of $()$-notation.

Using the recursion formula of \cite{GHIM} we previously
generated the explicit matrix formulations of the ``bases'' for any value of $\cal N$.  For 
the magic number cases where $1 \le {\cal N} \le 16$ leads to
\be\eqalign{
{\bm  {\bm {\cal P}}}_{1} ~&=~  {\bm {\mathbb{I}}}{}_{2} ~~ ~,~~  \cr
  {\bm {\cal P}}_{2} ~&=~ {\bm \s}^1    \,  
 ~~~~ , } \label{N2dia} \ee 
for ${\cal N}$ = 2
\be \begin{array}{ccccccccc}
{\bm {\cal P}}_{1} &=&  \, {\bm {\mathbb{I}}}{}_{2} & \otimes & {\bm {\mathbb{I}}}
{}_{2} & &  &  & ~~~, \\
{\bm {\cal P}}_{2} &=& {\bm \s}^1 & \otimes &{\bm \s}^1  & &   & &  ~~~, \\
{\bm {\cal P}}_{3} &=& {\bm \s}^1 & \otimes & {\bm {\mathbb{I}}}{}_{2}  & &   & &  ~~~, \\
{\bm {\cal P}}_{4} &=& {\bm {\mathbb{I}}}{}_{2}  & \otimes &{\bm \s}^1  & &   & &    ~~~, \\ 
\end{array} \label{N4dia}  \ee
for ${\cal N}$ = 4,
\be  \begin{array}{cccccccccccc}
{\bm {\cal P}}_{1} &=& \,  {\bm {\mathbb{I}}}{}_{2}\ & \otimes & {\bm {\mathbb{I}}}{}_{2 
} & \otimes & {\bm {\mathbb{I}}}{}_{2}  & & & & &~~~, ~~~ \\
{\bm {\cal P}}_{2} &=&  {\bm {\mathbb{I}}}{}_{2} & \otimes &{\bm {\mathbb{I}}}{}_{2} & 
\otimes &{\bm \s}^1  & &   & & &~~~, ~~~ \\
{\bm {\cal P}}_{3} &=& {\bm {\mathbb{I}}}{}_{2}  & \otimes &{\bm \s}^1 & \otimes & {\bm 
{\rm I}}{}_{2} & &    & & &~~~, ~~~ \\
{\bm {\cal P}}_{4} &=&  {\bm {\mathbb{I}}}{}_{2} & \otimes &{\bm \s}^1 & \otimes &{\bm 
\s}^1  & &   & &  &~~~, ~~~ \\
{\bm {\cal P}}_{5} &=& {\bm \s}^1 & \otimes &{\bm \s}^1 & \otimes & {\bm {\mathbb{I}}}{}_{2 
}  & &   & &  &~~~, ~~~ \\
{\bm {\cal P}}_{6} &=& {\bm \s}^1 & \otimes & {\bm {\mathbb{I}}}{}_{2}  & \otimes & {\bm 
\s}^1 & &    & & &~~~, ~~~ \\
{\bm {\cal P}}_{7} &=& {\bm \s}^1 & \otimes & {\bm {\mathbb{I}}}{}_{2} & \otimes &{\bm 
{\rm I}}{}_{2}  & &   & &  &~~~, ~~~ \\
{\bm {\cal P}}_{8} &=& {\bm \s}^1 & \otimes &{\bm \s}^1 & \otimes &{\bm \s}^1 
& &    & & &~~~, ~~~ \\
\end{array}
\label{N8diaX} \ee
for ${\cal N}$ = 8,
\be \begin{array}{ccccccccccccc}
 {\bm {\cal P}}{}_1&=& {\bm {\mathbb{I}}}{}_{2}& \otimes & {\bm {\mathbb{I}}}{}_{2} & \otimes & 
{\bm {\mathbb{I}}}{}_{2}&\otimes & {\bm {\mathbb{I}}}{}_{2}& & &  &  ~~~, \\
  {\bm {\cal P}}{}_2&=&   {\bm \s}^1  & \otimes & {\bm {\mathbb{I}}}{}_{2}
 & \otimes &  {\bm {\mathbb{I}}}{}_{2} &  \otimes &
 {\bm {\mathbb{I}}}{}_{2} &  &    &  & ~~~,  \\
 {\bm {\cal P}}{}_3&=& {\bm \s}^1 & \otimes & {\bm {\mathbb{I}}}{}_{2} & \otimes &
{\bm {\mathbb{I}}}{}_{2}   &  \otimes &  {\bm \s}^1 & &    & &  ~~~,  \\
 {\bm {\cal P}}{}_4 &=& {\bm \s}^1 & \otimes &{\bm {\mathbb{I}}}{}_{2}  & \otimes & {\bm \s}^1 &
 \otimes & {\bm {\mathbb{I}}}{}_{2}
 & &    & &  ~~~,  \\
 {\bm {\cal P}}{}_5&=& {\bm \s}^1 & \otimes & {\bm {\mathbb{I}}}{}_{2} & \otimes & {\bm \s}^1 &  
\otimes & {\bm \s}^1 & &    & & ~~~, \\
 {\bm {\cal P}}{}_6&=& {\bm \s}^1& \otimes & {\bm \s}^1 & \otimes & {\bm \s}^1 & 
\otimes & {\bm {\mathbb{I}}}{}_{2} & &    & &  ~~~,  \\
 {\bm {\cal P}}{}_7 &=& {\bm \s}^1 & \otimes &{\bm \s}^1 & \otimes &
{\bm {\mathbb{I}}}{}_{2} & \otimes & {\bm \s}^1& &    &  &  ~~~, \\
 {\bm {\cal P}}{}_8&=& {\bm \s}^1 & \otimes &{\bm \s}^1 & \otimes &
{\bm {\mathbb{I}}}{}_{2} & \otimes & {\bm {\mathbb{I}}}{}_{2} & &    & &  ~~~, \\
 {\bm {\cal P}}{}_9&=& {\bm \s}^1 & \otimes & {\bm \s}^1 & 
\otimes & {\bm \s}^1 & \otimes &{\bm \s}^1
&  &    & &  ~~~,
\end{array} \label{N9dia}  \ee
for ${\cal N}$ = 9, 
\be
\begin{array}{cccccccccccccccc} 
 {\bm {\cal P}}_{1} &=&{\bm {\mathbb{I}}}{}_{2} &\otimes & {\bm {\mathbb{I}}}{}_{2} & \otimes & {\bm {\mathbb{I}}}{}_{2}& \otimes & 
{\bm {\mathbb{I}}}{}_{2} & \otimes &{\bm {\mathbb{I}}}{}_{2} & &  & & , \\
 {\bm {\cal P}}_{2} &=& {\bm \s}^1 &\otimes & {\bm {\mathbb{I}}}{}_{2} & \otimes & {\bm {\mathbb{I}}}{}_{2}& \otimes & 
{\bm {\mathbb{I}}}{}_{2} & \otimes &{\bm {\mathbb{I}}}{}_{2} & &  & & , \\
 {\bm {\cal P}}_{3} &=& {\bm {\mathbb{I}}}{}_{2}  &\otimes &{\bm \s}^1 &\otimes& {\bm {\mathbb{I}}}{}_{2}& \otimes &{\bm {\mathbb{I}}}{}_{2}
&\otimes & {\bm {\mathbb{I}}}{}_{2}& &   & &,   \\
 {\bm {\cal P}}_{4}  &=&  {\bm \s}^1 &\otimes & {\bm {\mathbb{I}}}{}_{2} &\otimes &
{\bm {\mathbb{I}}}{}_{2} & \otimes &{\bm {\mathbb{I}}}{}_{2}  & \otimes &{\bm \s}^1
& &   & &,  \\
 {\bm {\cal P}}_{5} &=&  {\bm \s}^1& \otimes & {\bm {\mathbb{I}}}{}_{2} &\otimes &
{\bm {\mathbb{I}}}{}_{2}  & \otimes &{\bm \s}^1 & \otimes & {\bm {\mathbb{I}}}{}_{2} 
& &   & &,  \\
 {\bm {\cal P}}_{6} &=& {\bm \s}^1 &\otimes & {\bm {\mathbb{I}}}{}_{2} & \otimes &
{\bm {\mathbb{I}}}{}_{2} & \otimes &{\bm \s}^1 & \otimes &{\bm \s}^1 
& &   & & , \\
 {\bm {\cal P}}_{7} &=&  {\bm \s}^1 & \otimes & {\bm {\mathbb{I}}}{}_{2} & \otimes &
{\bm \s}^1 & \otimes &{\bm \s}^1 & \otimes & {\bm {\mathbb{I}}}{}_{2} 
&  &   & & , \\
 {\bm {\cal P}}_{8} &=&  {\bm \s}^1&\otimes & {\bm {\mathbb{I}}}{}_{2} & \otimes & 
{\bm \s}^1 & \otimes & {\bm {\mathbb{I}}}{}_{2}  & \otimes & {\bm \s}^1
& &   & & , \\
 {\bm {\cal P}}_{9} &=&  {\bm \s}^1 &\otimes & {\bm {\mathbb{I}}}{}_{2} & \otimes &
{\bm \s}^1 & \otimes & {\bm {\mathbb{I}}}{}_{2} & \otimes &{\bm {\mathbb{I}}}{}_{2}  
& &   & & , \\
 {\bm {\cal P}}_{10} &=&  {\bm \s}^1 &\otimes & {\bm {\mathbb{I}}}{}_{2} & \otimes & 
{\bm \s}^1 & \otimes &{\bm \s}^1 & \otimes &{\bm \s}^1
& &   &\, & , \\
 \end{array} \label{N10dia} \ee
for ${\cal N}$ = 10, 
\be
\begin{array}{ccccccccccccccc}
 {\bm {\cal P}}_{1}&=& {\bm {\mathbb{I}}}{}_{2} & \otimes &{\bm {\mathbb{I}}}{}_{2}& \otimes&{\bm {\mathbb{I}}}
{}_{2}& \otimes &{\bm {\mathbb{I}}}{}_{2} & \otimes & {\bm {\mathbb{I}}}{}_{2} & \otimes & {\bm 
{\rm I}}{}_{2} &  &    ~~,~  \\
 {\bm {\cal P}}_2& =  & {\bm {\s}} ^1 & \otimes &{\bm {\mathbb{I}}}{}_{2}& \otimes & {\bm {\mathbb{I}}}{}_{2} 
& \otimes &{\bm {\mathbb{I}}}{}_{2} & \otimes & {\bm {\mathbb{I}}}{}_{2} & \otimes & {\bm {\mathbb{I}}}{}_{2 
\times 2} &  & \, \,~~ ,~~\\ 
 {\bm {\cal P}}_3 & = &  {\bm {\mathbb{I}}}{}_{2} & \otimes & {\bm {\s}}^1 & \otimes & {\bm {\s}} ^1 & \otimes & {\bm {\mathbb{I}}}{}_{
2} & \otimes & {\bm {\mathbb{I}}}{}_{2} & \otimes & {\bm {\mathbb{I}}}{}_{2} &  &  \, 
~~~ , ~~\\ 
 {\bm {\cal P}}_4 & = &  {\bm {\mathbb{I}}}{}_{2} & \otimes & {\bm {\s}} ^1 & \otimes & {\bm {\mathbb{I}}}{}_{2} & \otimes & {\bm {\mathbb{I}}}{}_{2} & \otimes  & {\bm {\mathbb{I}}}{}_{2} & \otimes & {\bm 
 {\rm I}}{}_{2} &  &   \, ~~~ , ~~\\ 
 {\bm {\cal P}}_5&= & {\bm {\mathbb{I}}}{}_{2} & \otimes &{\bm {\mathbb{I}}}{}_{2}& \otimes&{\bm {\s}} ^1 
 &\otimes &{\bm {\mathbb{I}}}{}_{2} & \otimes & {\bm {\mathbb{I}}}{}_{2} & \otimes & {\bm {\mathbb{I}}}{}_{2 
 \times 2} &  &   \, ~~~ , ~~\\ 
 {\bm {\cal P}}_{6}&= & {\bm {\s}}^1 & \otimes &{\bm {\mathbb{I}}}{}_{2}& \otimes&
{\bm {\mathbb{I}}}{}_{2}& \otimes &  {\bm {\mathbb{I}}}{}_{2}
 & \otimes & {\bm {\mathbb{I}}}{}_{2} & \otimes & {\bm {\s}} ^1 &  &  \,    ~~~ , ~~\\ 
 {\bm {\cal P}}_7&= & {\bm {\s}}^1 & \otimes &{\bm {\mathbb{I}}}{}_{2}& \otimes&
{\bm {\mathbb{I}}}{}_{2} &\otimes &{\bm 
{\s}}^3 & \otimes &  {\bm {\s}} ^1  & \otimes & {\bm {\mathbb{I}}}{}_{2} &  &   \, 
~~~ , ~~\\ 
 {\bm {\cal P}}_8&= & {\bm {\s}}^1 & \otimes &{\bm {\mathbb{I}}}{}_{2}& \otimes &{\bm {\mathbb{I}}}{}_{2} &
\otimes & {\bm {\mathbb{I}}}{}_{2} & \otimes & {\bm {\s}}^1 & \otimes &  {\bm {\s}} ^1 
 &  &   \, 
~~~ , ~~\\ 
 {\bm {\cal P}}_9&= & {\bm {\s}}^1 & \otimes &{\bm {\mathbb{I}}}{}_{2}& \otimes&{\bm {\mathbb{I}}}{}_{2}& 
\otimes &{\bm {\s}} ^1 & \otimes & {\bm {\mathbb{I}}}{}_{2} & \otimes & {\bm {\mathbb{I}}}{}_{2} &  &   \, 
~~~ , ~~\\ 
 {\bm {\cal P}}_{10}&= & {\bm {\s}}^1 & \otimes &{\bm {\mathbb{I}}}{}_{2}& \otimes&
{\bm {\mathbb{I}}}{}_{2} 
& \otimes & {\bm {\s}} ^1 & \otimes &
{\bm {\mathbb{I}}}{}_{2} 
& \otimes &
 {\bm {\s}}^1 &  &   \,   ~~~ , ~~\\ 
 {\bm {\cal P}}_{11}&= & {\bm {\s}}^1 & \otimes &{\bm {\mathbb{I}}}{}_{2}& \otimes&
{\bm {\mathbb{I}}}{}_{2} 
& \otimes & {\bm {\s}} ^1 & \otimes &
{\bm {\mathbb{I}}}{}_{2} 
& \otimes &
 {\bm {\mathbb{I}}}{}_{2} &  &   \,   ~~~ , ~~\\ 
 {\bm {\cal P}}_{12}&= & {\bm {\s}}^1 & \otimes &{\bm {\mathbb{I}}}{}_{2}& \otimes&
{\bm {\mathbb{I}}}{}_{2}& \otimes &{\bm 
{\s}} ^1 & \otimes & {\bm {\s}} ^1 & \otimes & {\bm {\s}} ^1 &  &  \,    ~~~ , ~~\\ 
\end{array}
\label{N12dia} 
\ee
for ${\cal N}$ = 12, and 
\be
\begin{array}{ccccccccccccccccrc}
 {\bm {\cal P}}_{1} &=& {\bm {\mathbb{I}}}{}_{2} & \otimes & {\bm {\mathbb{I}}}{}_{2} & \otimes & {\bm {\mathbb{I}}}{}_{2} & \otimes & {\bm {\mathbb{I}}}{}_{2}&\otimes& {\bm {\mathbb{I}}}{}_{2}&\otimes &{\bm {\mathbb{I}}}{}_{2} &\otimes &{\bm {\mathbb{I}}}{}_{2}  & &    &, \\
 {\bm {\cal P}}_2 & =&  {\bm {\s}} ^1 &\otimes & {\bm {\mathbb{I}}}{}_{2}& \otimes & {\bm {\mathbb{I}}}{}_{2} & \otimes 
& {\bm {\mathbb{I}}}{}_{2} & \otimes & {\bm {\mathbb{I}}}{}_{2} & \otimes & {\bm {\mathbb{I}}}{}_{2} & \otimes & 
{\bm {\mathbb{I}}}{}_{2} &   &    \, &, \\
 {\bm {\cal P}}_3 & =&  {\bm {\mathbb{I}}}{}_{2} &\otimes & {\bm {\mathbb{I}}}{}_{2} & \otimes &  {\bm {\mathbb{I}}}{}_{2} & \otimes & {\bm {\s}} ^1 & \otimes& {\bm {\mathbb{I}}}{}_{2} & \otimes & {\bm {\mathbb{I}}}{}_{2} & 
\otimes &{\bm {\mathbb{I}}}{}_{2}  & &     \,  &, \\
 {\bm {\cal P}}_4 & =&  {\bm {\mathbb{I}}}{}_{2}&\otimes& {\bm {\mathbb{I}}}{}_{2} & \otimes& {\bm {\s}} ^1 & \otimes 
 & {\bm {\mathbb{I}}}{}_{2} &\otimes & {\bm {\mathbb{I}}}{}_{2} & \otimes &{\bm {\mathbb{I}}}{}_{2}& \otimes &{\bm 
 {\rm I}}{}_{2}  & &     \,  &, \\
 {\bm {\cal P}}_5 &= & {\bm {\mathbb{I}}}{}_{2} &\otimes &{\bm {\mathbb{I}}}{}_{2}& \otimes &{\bm {\s}}^1 &\otimes 
 &{\bm {\s}} ^1 &\otimes & {\bm {\mathbb{I}}}{}_{2}& \otimes & {\bm {\mathbb{I}}}{}_{2} & \otimes & {\bm {\mathbb{I}}}{}_{2}  & &   \,  &, \\
 {\bm {\cal P}}_6 &=&  {\bm {\mathbb{I}}}{}_{2} &\otimes &{\bm {\s}}^1 &\otimes& {\bm {\s}} ^1 & \otimes &
 {\bm {\mathbb{I}}}{}_{2} &\otimes &{\bm {\mathbb{I}}}{}_{2} &\otimes & {\bm {\mathbb{I}}}{}_{2}&\otimes & {\bm 
 {\rm I}}{}_{2} & &   \,  &, \\
 {\bm {\cal P}}_7 &= & {\bm {\mathbb{I}}}{}_{2} &\otimes &{\bm {\s}} ^1 &\otimes& {\bm {\mathbb{I}}}{}_{2}& \otimes 
 &{\bm {\s}}^1 &\otimes &{\bm {\mathbb{I}}}{}_{2}&\otimes& {\bm {\mathbb{I}}}{}_{2} & \otimes &{\bm {\mathbb{I}}}
 {}_{2} & &   \,  &, \\
 {\bm {\cal P}}_8 &= & {\bm {\mathbb{I}}}{}_{2} &\otimes &{\bm {\s}} ^1 &\otimes& {\bm {\mathbb{I}}}{}_{2}& \otimes 
 &{\bm {\mathbb{I}}}{}_{2} &\otimes &{\bm {\mathbb{I}}}{}_{2}&\otimes& {\bm {\mathbb{I}}}{}_{2} & \otimes &{\bm 
 {\rm I}}{}_{2} & &   \,  &, \\
{\bm {\cal P}}_9 &=& {\bm {\mathbb{I}}}{}_{2}&\otimes&{\bm {\s}} ^1 &\otimes &{\bm {\s}} ^1 &\otimes &{\bm {\s}}^1 &
\otimes & {\bm {\mathbb{I}}}{}_{2}  &\otimes& {\bm {\mathbb{I}}}{}_{2}&\otimes &{\bm {\mathbb{I}}}{}_{2}  & &  \,  
 &, \\
 {\bm {\cal P}}_{10} &=&  {\bm {\s}}^1& \otimes & {\bm {\mathbb{I}}}{}_{2}& \otimes& {\bm {\mathbb{I}}}{}_{2}& 
 \otimes &{\bm {\mathbb{I}}}{}_{2}&\otimes&{\bm {\mathbb{I}}}{}_{2}&\otimes&{\bm {\mathbb{I}}}{}_{2}&\otimes 
 &{\bm {\s}}^1 & &  \,  &,\\
 {\bm {\cal P}}_{11} &= & {\bm {\s}}^1 &\otimes&{\bm {\mathbb{I}}}{}_{2}&\otimes&{\bm {\mathbb{I}}}{}_{2}& \otimes 
&{\bm {\mathbb{I}}}{}_{2}&\otimes &{\bm {\mathbb{I}}}{}_{2} &\otimes &{\bm {\s}} ^1 &\otimes&{\bm {\mathbb{I}}}
{}_{2}& &  \,  &,\\
 {\bm {\cal P}}_{12} &=&  {\bm {\s}}^1 &\otimes&{\bm {\mathbb{I}}}{}_{2}&\otimes&{\bm {\mathbb{I}}}{}_{2}&\otimes 
&{\bm {\mathbb{I}}}{}_{2}&\otimes&{\bm {\mathbb{I}}}{}_{2}&\otimes&{\bm {\s}}^1&\otimes&{\bm {\s}} ^1& &  \,    &, \\
 {\bm {\cal P}}_{13} &= & {\bm {\s}}^1 &\otimes& {\bm {\mathbb{I}}}{}_{2}& \otimes &{\bm {\mathbb{I}}}{}_{2}&\otimes 
&{\bm {\mathbb{I}}}{}_{2}& \otimes &{\bm {\s}}^1 &\otimes& {\bm {\s}}^1 &\otimes& {\bm {\mathbb{I}}}{}_{2}& &  \,  &;\\
 {\bm {\cal P}}_{14} &=&  {\bm {\s}}^1 &\otimes& {\bm {\mathbb{I}}}{}_{2}& \otimes& {\bm {\mathbb{I}}}{}_{2}&\otimes 
&{\bm {\mathbb{I}}}{}_{2}&\otimes &{\bm {\s}} ^1 &\otimes&{\bm {\mathbb{I}}}{}_{2}& \otimes &{\bm {\s}}^1 & &   \,  &,\\
 {\bm {\cal P}}_{15} &=&  {\bm {\s}}^1 &\otimes& {\bm {\mathbb{I}}}{}_{2}&\otimes& {\bm {\mathbb{I}}}{}_{2}&\otimes 
&{\bm {\mathbb{I}}}{}_{2}& \otimes &{\bm {\s}}^1 &\otimes& {\bm {\mathbb{I}}}{}_{2}&\otimes &{\bm {\mathbb{I}}}
{}_{2} & &  \,  &,\\
 {\bm {\cal P}}_{16} &= & {\bm {\s}}^1 &\otimes&{\bm {\mathbb{I}}}{}_{2}&\otimes&{\bm {\mathbb{I}}}{}_{2}&\otimes
&{\bm {\mathbb{I}}}{}_{2} &\otimes &{\bm {\s}} ^1& \otimes &{\bm {\s}} ^1& \otimes &{\bm {\s}}^1 & &  \,   &,\\
\end{array}
\label{N16dia} \ee 
for ${\cal N}$ = 16. In these equations, the quantities
${\bm {\s}}^1$,${\bm {\s}}^2$,and ${\bm {\s}}^3$ refer to
the usual 2 $\times$ 2 Pauli matrices and ${\bm {\rm {I}}}{}_{2} $ refer to the identity matrix.

After these are converted into $\langle  \rangle$-notation, it is possible to calculate the correlators for any of these cases.  This is a topic for future study.  Since there exists a recursion formula
to derive the last eight of these from the first eight, it is
conceivable that it should be possible to derive the intra-$\cal N$-tet correlators, along with their eigenvalues, and traces, in complete generality.

Collectively the sets of permutation objects that appear in Eq.(\ref{N2dia}) to Eq.(\ref{N16dia})
have informally been given the names of ``diadems.''

\newpage 
\section{Conclusion}
\label{sec:CONs}
In this work, we have used structures that are intrinsic to the Coxeter
Group ${\mathbb{BC}}{}_{4}$ in order to discuss how their elements are
organized to provide representations of four-dimensional $\cal N$ = 1
SUSY.

It is of interest to note none of the eigenvalues for the inter-quartet 
correlators align with that of the intra-quartet correlators.  It is also true 
that none of the traces of the inter-quartet correlators vanish as in the case 
of the intra-quartet correlators.

As seen from the results in Tables 7-21 only three values of traces and four sets of eigenvalues 
occurred for the distributions of the various inter-quartet eigenvalues and traces.  These are 
shown in the table below.
$$
{
\begin{tabular}{|c|c|c|c|c|} \hline
\diagbox{$ {\rm {Eigenvalues}} $}{$ {\rm {Traces}} $} 
& {~}4  & {~}8 & {~}12    \\ \hline
{$(12,0,0,0) $} & {}-  & {}- &  {\,} \VV  \\ \hline
{$(12,-4,0,0) $} & {} \WW & {} \XX
&  {}-    \\ \hline
{$(12,-4,-4,0) $} & {}
\YY 
& {}- &  {\,}-   \\ \hline
{$(12,-8,0,0) $} & {} \ZZ  
& {}- &  {\,}-   \\ \hline
\end{tabular} }
\label{tab:E-T}
$$
\be{{~}}
\ee
\begin{center}
{{\bf Table 23:}  {Eigenvalue/Trace Distribution of Two-Point Correlator Values}}
 \end{center}
\noindent
These can be compared to the eigenvalues and traces of the intra-quartet correlators were
given by $(12,0,-4,-8) $
and 0 respectively.

If we calculate the ``length'' (by use of the usual Euclidean metric) of the eigenvectors associated with the intra-quartet correlators, it takes the value of $\sqrt{ 224}$.  This may be compared to the ``lengths'' found for the inter-quartet correlator eigenvectors. Working from top to bottom in the right most column of Table 23, the respective lengths are
$\sqrt{ 144}$, $\sqrt{ 160}$, $\sqrt{ 176}$, and
$\sqrt{ 208}$, respectively.

Thus, we conclude that the SUSY quartets in ${\mathbb {S}}{}_4 $ are
precisely the ones that lead to the maximum possible value of the lengths of the eigenvectors
for quartet correlators.

The use of the two-point quartet correlators reveal symmetries of how the SUSY quartets are
embedded in the permutahedron in other ways also.  One of the most amusing interpretations is
that the SUSY quartets are the solutions to a set of Sudoku puzzles!

For example, looking at the results in Equations (\ref{eq:Mtrx1}), (\ref{eq:Mtrx2}),  (\ref{eq:Mtrx3}),
(\ref{eq:Mtrx4}), (\ref{eq:Mtrx5}), and (\ref{eq:Mtrx6}), it is apparent that every column and every
row in these matrices sum to twelve. One is required to place zeros in every diagonal entry and then
follow the remaining rules stated under the first full paragraph under (\ref{eq:Mtrx1}).

This Sudoku-puzzle solution interpretation also follows for every table (and therefore every associated
matrix) in the list of results shown in Equations (\ref{eq:T1}) - (\ref{eq:T15})
where the sums of the columns and rows is the same. The rules in these 
cases start with the a ``source set'' of numbers 1, $\dots $, 5, 6, but become
considerably more baroque to state.

We believe the most powerful implication of the observations in this work is that the representation
theory for SUSY for all values of $\cal N $ can be interpreted a Sudoku puzzle where the diadems
set the start of rules.  This means an essential part of a comprehensive understanding of the representation
rules for SUSY can be studied solely and a mathematical problem of diadem embeddings into 
${\mathbb S}{}_d$...with no reference at all to QFT.

  \vspace{.05in}
 \begin{center}
\parbox{4in}{{\it ``Living is worthwhile if one can contribute in some small \\ $~~$ 
way to this endless chain of progress.'' \\ ${~}$ 
 ${~}$ 
\\ ${~}$ }\,\,-\,
Paul A.\ M.\ Dirac $~~~~~~~~~$}
 \parbox{4in}{
 $~~$}  
 \end{center}
		
\noindent
{\bf Acknowledgments}\\[.1in] \indent
This research is supported in part by the endowment of the Ford Foundation Professorship 
of Physics at Brown University and the Brown Theoretical Physics Center. Additional 
acknowledgment is given for their participation in the 2020 SSTPRS (Student Summer 
Theoretical Physics Research Session) program by Aleksander Cianciara, 
and Ren\' ee Kirk.  SJG wishes to
further acknowledge the
Mathematical Sciences Research Institute (MSRI) for hosting the
2019 African Diaspora Joint Mathematics Workshop (ADJOINT) program.  The lively discussions with the other members of the ``Adinkra Plaquette'' (Professors Caroline Klivans, Kevin Iga, and Vincent Rodgers)
played an important role in the research that ultimately resulted in exploration of the permutahedron.
We also acknowledge Delilah Gates for an observation about the symmetries of the correlators that spurred 
discussion of the Sudoku comparison. 
Finally, we acknowledge preliminary discussions with Ismail Elemengad.

\noindent
{\bf Added Note In Proof}\\[.1in] \indent
The use of the phrase `Sudoku Puzzle' in the title is not meant in exactitude as the actual rules
of Sudoku are more stringent than the patterns seen in the `correlators' tables, as noted for us by
Prof.\ Tristan H\" ubsch. 

\newpage
\appendix
\section{Permutations Expressed As Matrices \label{appen:Pmatrices}}

In this appendix we simply express the permutations associated with ${\mathbb{S}}{}_{4}$ as 4 $\times $ 4 matrices.
$$
() ~=~ 
\begin{bmatrix}
1 & 0 & 0 & 0\\
0 & 1 & 0 & 0\\
0 & 0 & 1 & 0\\
0 & 0 & 0 & 1 
\end{bmatrix} ~~~,~~~
(12) ~=~ 
\begin{bmatrix}
0 & 1 & 0 & 0\\
1 & 0 & 0 & 0\\
0 & 0 & 1 & 0\\
0 & 0 & 0 & 1 
\end{bmatrix} ~~~,~~~
(13) ~=~ 
\begin{bmatrix}
0 & 0 & 1 & 0\\
0 & 1 & 0 & 0\\
1 & 0 & 0 & 0\\
0 & 0 & 0 & 1 
\end{bmatrix}  ~~~,~~~
$$
$$
(14) ~=~ 
\begin{bmatrix}
0 & 0 & 0 & 1\\
0 & 1 & 0 & 0\\
0 & 0 & 1 & 0\\
1 & 0 & 0 & 0 
\end{bmatrix} ~~~, ~~~
(23) ~=~ 
\begin{bmatrix}
1 & 0 & 0 & 0\\
0 & 0 & 1 & 0\\
0 & 1 & 0 & 0\\
0 & 0 & 0 & 1 
\end{bmatrix}~~~,~~~
(24) ~=~ 
\begin{bmatrix}
1 & 0 & 0 & 0\\
0 & 0 & 0 & 1\\
0 & 0 & 1 & 0\\
0 & 1 & 0 & 0 
\end{bmatrix} ~~~,~~~
$$
$$
(34) ~=~ 
\begin{bmatrix}
1 & 0 & 0 & 0\\
0 & 1 & 0 & 0\\
0 & 0 & 0 & 1\\
0 & 0 & 1 & 0 
\end{bmatrix} ~~~,~~~
(123) ~=~ 
\begin{bmatrix}
0 & 1 & 0 & 0\\
0 & 0 & 1 & 0\\
1 & 0 & 0 & 0\\
0 & 0 & 0 & 1 
\end{bmatrix} ~~~,~~~
(124) ~=~ 
\begin{bmatrix}
0 & 1 & 0 & 0\\
0 & 0 & 0 & 1\\
0 & 0 & 1 & 0\\
1 & 0 & 0 & 0 
\end{bmatrix} ~~~,~~~
$$
$$
(132) ~=~ 
\begin{bmatrix}
0 & 0 & 1 & 0\\
1 & 0 & 0 & 0\\
0 & 1 & 0 & 0\\
0 & 0 & 0 & 1 
\end{bmatrix} ~~~,~~~
(134) ~=~ 
\begin{bmatrix}
0 & 0 & 1 & 0\\
0 & 1 & 0 & 0\\
0 & 0 & 0 & 1\\
1 & 0 & 0 & 0 
\end{bmatrix}~~~,~~~
(142) ~=~ 
\begin{bmatrix}
0 & 0 & 0 & 1\\
1 & 0 & 0 & 0\\
0 & 0 & 1 & 0\\
0 & 1 & 0 & 0
\end{bmatrix}  ~~~,~~~
$$
$$
(143) ~=~ 
\begin{bmatrix}
0 & 0 & 0 & 1\\
0 & 1 & 0 & 0\\
1 & 0 & 0 & 0\\
0 & 0 & 1 & 0
\end{bmatrix} ~~~,~~~
(234) ~=~ 
\begin{bmatrix}
1 & 0 & 0 & 0\\
0 & 0 & 1 & 0\\
0 & 0 & 0 & 1\\
0 & 1 & 0 & 0
\end{bmatrix} ~~~,~~~
(243) ~=~ 
\begin{bmatrix}
1 & 0 & 0 & 0\\
0 & 0 & 0 & 1\\
0 & 1 & 0 & 0\\
0 & 0 & 1 & 0
\end{bmatrix} ~~~,~~~
$$
$$
(1234) ~=~ 
\begin{bmatrix}
0 & 1 & 0 & 0\\
0 & 0 & 1 & 0\\
0 & 0 & 0 & 1\\
1 & 0 & 0 & 0
\end{bmatrix} ~~~,~~~
(1243) ~=~ 
\begin{bmatrix}
0 & 1 & 0 & 0\\
0 & 0 & 0 & 1\\
1 & 0 & 0 & 0\\
0 & 0 & 1 & 0 
\end{bmatrix}~~~,~~~
(1324) ~=~ 
\begin{bmatrix}
0 & 0 & 1 & 0\\
0 & 0 & 0 & 1\\
0 & 1 & 0 & 0\\
1 & 0 & 0 & 0
\end{bmatrix} ~~~,~~~
$$
$$
(1342) ~=~ 
\begin{bmatrix}
0 & 0 & 1 & 0\\
1 & 0 & 0 & 0\\
0 & 0 & 0 & 1\\
0 & 1 & 0 & 0
\end{bmatrix} ~~~,~~~
(1423) ~=~ 
\begin{bmatrix}
0 & 0 & 0 & 1\\
0 & 0 & 1 & 0\\
1 & 0 & 0 & 0\\
0 & 1 & 0 & 0
\end{bmatrix} ~~~, ~~~
(1432) ~=~ 
\begin{bmatrix}
0 & 0 & 0 & 1\\
1 & 0 & 0 & 0\\
0 & 1 & 0 & 0\\
0 & 0 & 1 & 0
\end{bmatrix} ~~~,~~~
$$
$$
(12)(34) ~=~
\begin{bmatrix}
0 & 1 & 0 & 0\\
1 & 0 & 0 & 0\\
0 & 0 & 0 & 1\\
0 & 0 & 1 & 0
\end{bmatrix} ~~~,~~~
(13)(24) ~=~
\begin{bmatrix}
0 & 0 & 1 & 0\\
0 & 0 & 0 & 1\\
1 & 0 & 0 & 0\\
0 & 1 & 0 & 0
\end{bmatrix} ~~~,~~~
(14)(32) ~=~
\begin{bmatrix}
0 & 0 & 0 & 1\\
0 & 0 & 1 & 0\\
0 & 1 & 0 & 0\\
1 & 0 & 0 & 0
\end{bmatrix} ~~~. 
$$

\newpage
\section{Eigenvalue Equivalence Classes}\label{sec:classes}

\begin{table}[htp!]
    \centering
    \begin{tabular}{|c|c|c|}
    \hline
       Eigenvalues  & \# of Matrices & Trace \\\hline\hline
        {12, 8, 4, 0} & 144 & 24 \\\hline
        {12, 8, 0, 0} & 288 & 20 \\\hline
        {12, 4, 0, 0} & 288 & 16 \\\hline
        {12, 8, -4, 0} & 144  & 16 \\\hline
         {12, 0, 0, 0} & 1,152 & 12 \\\hline
         ${12, -4\sqrt{2}, 4\sqrt{2}, 0}$  & 288 & 12 \\\hline
        ${12, -4i\sqrt{2}, 4i\sqrt{2}, 0}$  & 288 & 12 \\\hline
        {12, -4, 0, 0} &288  & 8 \\\hline
         {12, -8, 4, 0} & 144 &  8\\\hline
          {12, -8, 0, 0} & 288 & 4 \\\hline
           {12, -8, -4, 0} & 144 & 0 \\\hline
    \end{tabular}
    \label{tab:sum_class1}
\end{table}
\begin{center}
{{\bf Table 24:} Eigenvalue Equivalent Classes of 3,456 Distance Matrices in Class (1)}
 \end{center} 

\begin{table}[htp!]
    \centering
    \begin{tabular}{|c|c|c|}
    \hline
       Eigenvalues  & \# of Matrices & Trace \\\hline\hline
        {12, 4, 0, 0} & 1,152 & 16 \\\hline
         {12, 0, 0, 0} & 4,608 & 12 \\\hline
        {12, -4, 0, 0} & 1,152 & 8 \\\hline
    \end{tabular}
    \label{tab:sum_class2}
\end{table}
\begin{center}
{{\bf Table 25:} Eigenvalue Equivalent Classes of 6,912 Distance Matrices in Class (2.1)}
 \end{center}

\begin{table}[htp!]
    \centering
    \begin{tabular}{|c|c|c|}
    \hline
       Eigenvalues  & \# of Matrices & Trace \\\hline\hline
        {12, 8, 0, 0} & 576 & 20 \\\hline
         {12, 0, 0, 0} & 2,304 & 12 \\\hline
          {12, -8, 0, 0} & 576 & 4 \\\hline
    \end{tabular}
    \label{tab:sum_class3}
\end{table}
\begin{center}
{{\bf Table 26:} Eigenvalue Equivalent Classes of 3,456 Distance Matrices in Class (2.2)}
 \end{center}
\newpage 
\begin{table}[htp!]
    \centering
    \begin{tabular}{|c|c|c|}
    \hline
       Eigenvalues  & \# of Matrices & Trace \\\hline\hline
         {12, 0, 0, 0} & 3,456 & 12 \\\hline
    \end{tabular}
    \label{tab:sum_class4}
\end{table}
\begin{center}
{{\bf Table 27:} Eigenvalue Equivalent Classes of 3,456 Distance Matrices in Class (2.3)}
 \end{center}

\begin{table}[htp!]
    \centering
    \begin{tabular}{|c|c|c|}
    \hline
       Eigenvalues  & \# of Matrices & Trace \\\hline\hline
        {12, 4, 4,0} & 144 & 20 \\\hline
        {12, 4, 0, 0} & 576 & 16 \\\hline
         {12, 0, 0, 0} & 1,152 & 12 \\\hline
         {12, -4, 4, 0} & 576 & 12 \\\hline
 ${12, 4i, -4i, 0}$ & 288 & 12 \\\hline
        {12, -4, 0, 0} & 576 & 8 \\\hline
 {12, -4, -4, 0} & 144 & 4 \\\hline
    \end{tabular}
    \label{tab:sum_class5}
\end{table}
\begin{center}
{{\bf Table 28:} Eigenvalue Equivalent Classes of 3,456 Distance Matrices in Class (2.4)}
\end{center}

\newpage

\end{document}